\documentclass[10pt,english,aps, prd, twocolumn,floatfix]{revtex4}
\usepackage[T1]{fontenc}
\usepackage[latin9]{inputenc}
\usepackage{refstyle}
\usepackage{float}
\usepackage{textcomp}
\usepackage{amsmath}
\usepackage{amssymb}
\usepackage{graphicx}
\usepackage{esint}

\makeatletter

\AtBeginDocument{\providecommand\figref[1]{\ref{fig:#1}}}
\AtBeginDocument{\providecommand\eqref[1]{\ref{eq:#1}}}
\AtBeginDocument{\providecommand\secref[1]{\ref{sec:#1}}}
\AtBeginDocument{\providecommand\subsecref[1]{\ref{subsec:#1}}}
\AtBeginDocument{\providecommand\tabref[1]{\ref{tab:#1}}}
\providecommand{\tabularnewline}{\\}
\RS@ifundefined{subsecref}
  {\newref{subsec}{name = \RSsectxt}}
  {}
\RS@ifundefined{thmref}
  {\def\RSthmtxt{theorem~}\newref{thm}{name = \RSthmtxt}}
  {}
\RS@ifundefined{lemref}
  {\def\RSlemtxt{lemma~}\newref{lem}{name = \RSlemtxt}}
  {}

\@ifundefined{textcolor}{}
{%
 \definecolor{BLACK}{gray}{0}
 \definecolor{WHITE}{gray}{1}
 \definecolor{RED}{rgb}{1,0,0}
 \definecolor{GREEN}{rgb}{0,1,0}
 \definecolor{BLUE}{rgb}{0,0,1}
 \definecolor{CYAN}{cmyk}{1,0,0,0}
 \definecolor{MAGENTA}{cmyk}{0,1,0,0}
 \definecolor{YELLOW}{cmyk}{0,0,1,0}
}

\usepackage{lmodern}
\usepackage[T1]{fontenc}
\usepackage{prettyref}
\usepackage{hyperref}
\usepackage{textgreek}
\usepackage{inputenc}
\usepackage{slashed}

\makeatother

\usepackage{babel}
\begin{document}

\title{The impact of the quantized transverse motion on radiation emission
in a Dirac harmonic oscillator}

\author{Tobias N. Wistisen and Antonino Di Piazza}

\affiliation{Max-Planck-Institut f{\"u}r Kernphysik, Saupfercheckweg 1, D-69117, Germany}
\begin{abstract}
We investigate the radiation emitted by an ultrarelativistic electron
traveling in a 1-dimensional parabolic potential. Having in mind
a simplified model for beamstrahlung, we consider the realistic case
of the electron motion being highly directional, with the transverse
momentum being much smaller than the longitudinal one. In this case we
can find solutions of the Dirac equation and we calculate exactly
the radiation emission using first-order perturbation theory. 
We compare the results obtained to that obtained via the semi-classical 
method of Baier and Katkov which includes quantum effects due to
photon recoil in the radiation emission but ignores the quantum
nature of the electron motion. On the one hand, we confirm a
prediction of the semi-classical method that the emission spectrum
should coincide with that in the case of a linearly polarized
monochromatic wave. On the other hand, however, we find that the
semi-classical method does not yield the exact result when the 
quantum number describing the transverse motion becomes
small. In this way, we address quantitatively the problem of
the limits of validity of the semi-classical method, which is
known, generally speaking, to be applicable for large quantum
numbers. Finally, we also discuss which beam conditions
would be necessary to enter the studied regime where also the 
motion of the particles must be considered quantum mechanically 
to yield the correct spectrum.
\end{abstract}
\maketitle

\section{Introduction}

Strong-field QED is the study of electromagnetic
phenomena in the presence of background electromagnetic fields,
whose strength approaches a critical limit, called the ``Schwinger limit'' \cite{Mitter_1975,Ritus_1985,Ehlotzky_2009,Di_Piazza_2012,Roshchupkin_2012}. 
In this limit phenomena that are of a purely quantum mechanical nature arise, 
such as pair production and vacuum birefringence \cite{PhysRev.135.B1279,PhysRevD.89.125003,Heinzl2006318,1751-8121-40-5-F01,PhysRevD.88.053009,PhysRevLett.119.250403,Narozhny_2000,Roshchupkin_2001,Heinzl_2010b,Mueller_2011b,Titov_2012,Nousch_2012,Krajewska_2013b,Jansen_2013,Augustin_2014,Di_Piazza_2008,Meuren_2011,Di_Piazza_2013,PhysRevD.89.125003,Dinu_2014b,Gies_2014},
and quantum effects in radiation emission become essential \cite{Ritus_1985,Baie98,PhysRevLett.110.070402,2017arXiv170808276D,PhysRevLett.120.044802,fuchs2015anomalous,PhysRevD.92.045045,RAICHER201576,PhysRevD.86.072001,PhysRevLett.76.3116,Narozhny_2000,Boca_2009,Harvey_2009,Heinzl_2009,Mackenroth_2010,Mackenroth_2011,Seipt_2011,Krajewska_2012,Seipt_2013,Dinu_2013,Nedoreshta_2013,Krajewska_2014,Wistisen_2014,Angioi_2016,Di_Piazza_2017b,Boca_2009,PhysRevA.91.033415}.
Most of the mentioned studies consider a plane-wave as a background field,
having in mind processes occurring in the presence of a strong laser field.
In this respect, there is also a growing interest in finding out how the
basic strong-field QED processes mentioned above are altered in the presence 
of a laser beams tightly focused in space other than in time \cite{Di_Piazza_2014,Di_Piazza_2015,Di_Piazza_2016,Heinzl_2016,Di_Piazza_2017b,Heinzl_2017,Heinzl_2017b}.
In this paper, however, our focus is not on laser fields, but on different methods
of calculating radiation emission, in particular a comparison between
a fully quantum calculation compared to a semi-classical method, which
can also be used for the case of lasers \cite{PhysRevD.89.125003}.

The semi-classical operator method developed by Baier and Katkov in
1968 \cite{baier1968processes} is a powerful method to calculate
radiation emission and the probabilities of other quantum processes.
Quantum effects such as spin and recoil during emission are included
in the method but the motion of the charged particle is considered
as classical, i.e., along a trajectory. Thus, in order to calculate
the quantum observables, only the particle's trajectory is needed, which can be found
numerically in an arbitrary field configuration. Using this method to calculate nonlinear 
Compton scattering in more complex field configurations was the focus 
of \cite{Wistisen_2014}. Now, finding the wave-function of an
electron in any given field configuration is in general an impossible
task and it is therefore prudent to ask exactly when the method of Baier 
and Katkov is applicable. This is of course discussed by the authors 
themselves and the mentioned conditions are that the particle should 
be ultra-relativistic and that the commutator among the operators corresponding
to different velocity components should be negligibly small, in the sense that \cite{Baie98}
\begin{equation}
\frac{\left|\left\langle \left[\hat{\Pi}^{\mu},\hat{\Pi}^{\nu}\right]\right\rangle \right|}{\varepsilon^{2}}=\frac{e\left|F^{\mu\nu}(x)\right|}{\varepsilon^{2}}\ll1,\label{eq:commutator}
\end{equation}
where $\hat{\Pi}^{\mu}=\hat{p}^{\mu}+eA^{\mu}(x)$, $\hat{p}^{\mu}$
is the four-momentum operator, $e>0$ is the elementary charge, 
$A^{\mu}(x)=(\varphi(x),\boldsymbol{A}(x))$ is the four-vector potential of the external field, 
$F^{\mu\nu}(x)$ is the electromagnetic field tensor and $\varepsilon$ 
is the particle energy. This condition is, indeed, fulfilled for any 
currently imaginable electromagnetic field. For the well-known exact 
solutions of the Dirac equation in the field configurations of a 
plane wave \cite{Ritus_1985}, the semi-classical operator method yields
exactly the same result as the full quantum calculation. In the final
step of the derivation of the method in \cite{baier1968processes},
it is stated that since the unfolding of a certain operator has been
performed, the expectation value of this operator can be replaced
by its corresponding classical value. This, however, may not always
be allowed, even when the previously mentioned conditions 
are fulfilled. The field configuration studied in the present paper is 
an example of this. In the book \cite{Baie98} by the same authors, 
an additional condition has been added, that one can replace the expectation 
value of the operator with the classical value when the quantum state of the
electron is characterized by large quantum numbers. This is in line with 
Bohr's correspondence principle. In the present paper we investigate
exactly this aspect of the method of Baier and Katkov: the method reproduces 
the correct quantum result when the quantum numbers describing the motion 
are large. In \cite{Baie98} the semi-classical method has, naturally, also 
been employed to study the radiation from the relativistic harmonic oscillator, 
which is in essence the problem studied here. In this paper we also include 
a magnetic field such that the field also can be employed as a simplified model 
of ``beamstrahlung'', i.e. the radiation emitted when high-energy dense charged
bunches collide. In our solution we can turn off the magnetic
field component and thus also obtain the results of the harmonic oscillator
as in \cite{Baie98}. In fact, the only effect of the magnetic field
is to effectively make the oscillator twice as strong. Usually, in
future linear colliders, the colliding bunches are of identical shape
and oppositely charged, i.e. an electron bunch colliding with a positron bunch.
In this case during the collision, the field from one bunch will alter
the shape of the other bunch and vice versa. The full problem is therefore
multi-particle, making it complicated to fully solve it quantum mechanically.
However the classical motion can be solved in this case, and therefore
the semi-classical method of Baier and Katkov can be applied. In this
paper, however, the main interest is not to make a precise analysis
of beamstrahlung, but to investigate when and why the semi-classical
approach breaks down. We therefore consider the case where a single electron
interacts with the field of a positron bunch as in this way the
positron bunch can be assumed not to change shape during the collision. 
This solution would still be valid
if one studies the collision of a low-density bunch with a high-density 
bunch, such that the low-density bunch has only negligible effect on 
the dense one.

As shown in \cite{Baie98}, the result of the semi-classical operator
method applied to the one-dimensional oscillator problem yields simply 
the spectrum obtained in the case of nonlinear Compton scattering in a linearly
polarized monochromatic plane wave as found in, e.g., \cite{Ritus_1985}.
As we shall see, the correct calculation will deviate from this result when
the quantum number of the discretized transverse motion becomes small.
However, since this comparison is an important point, below we will also apply the semi-classical operator method to this problem \cite{Baie98}. 
In \secref{Model-of-the} we will first make some considerations on the 
electromagnetic field generated from the relativistic positron bunch and indicate 
how one arrives at the parabolic potential approximation. In \secref{Classical-motion}
we will gain an intuition of the problem and find an approximated analytical solution
of the classical equations of motion of the problem, enabling us to
apply the semi-classical method of Baier and Katkov in \secref{Baier-method}.
In \secref{Solution-of-Dirac-1} we find the approximate wave-functions
for the problem at hand and in \secref{rademit} we use these wave-functions
to calculate the transition matrix element of the single-photon radiation emission.
In \secref{discuss} we do a side-by-side comparison of the power spectra
obtained using the two methods of calculation and discuss the different
regimes of radiation emission which arise. Finally in \secref{Conclusion}
we draw the main conclusions of the paper. 

We use units where $\hbar=c=1$, $\alpha=e^{2}$ and the Feynman slash notation 
such that $\slashed a=a_{\mu}\gamma^{\mu}$, where $\gamma^{\mu}$ are the Dirac 
gamma matrices and $a^{\mu}$ an arbitrary four-vector. We adopt the metric tensor 
$\eta^{\mu\nu}=\text{diag}(+1,-1,-1,-1)$.

\section{Model of the field\label{sec:Model-of-the}}

\begin{figure}[t]
\includegraphics[width=1\columnwidth]{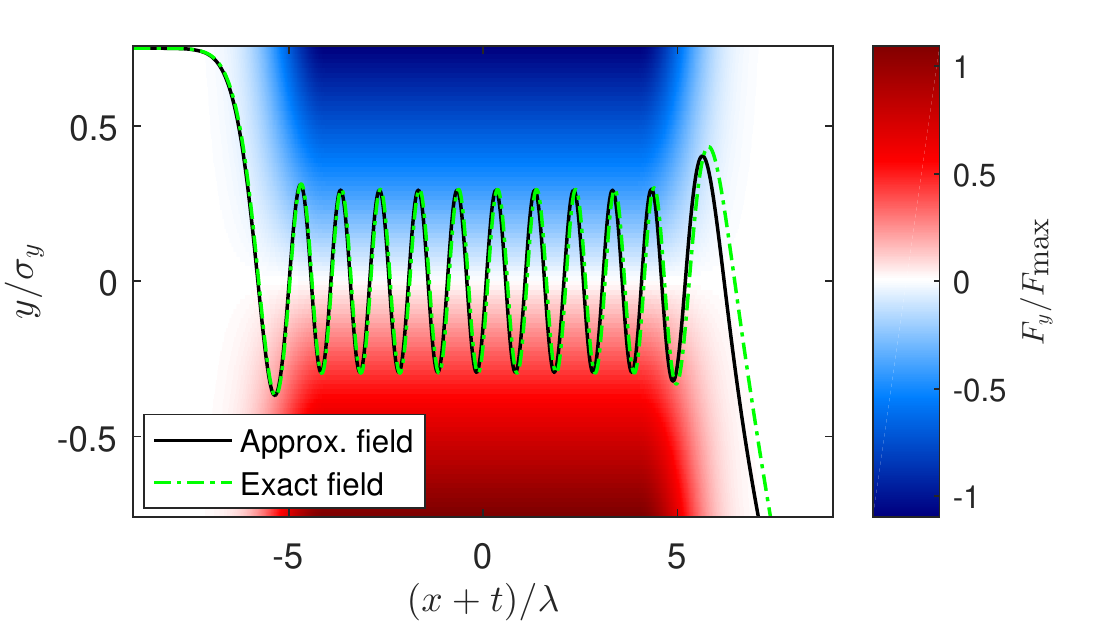}
\caption{The $y$-component $F_{y}$ of the force exerted by a positron
bunch propagating along the positive $x$-direction and lying on the $x$-$z$ plane
as function of position and time. Here, $F_{\text{max}}$ is the magnitude of the
force at $x+t=0$ and $y/\Sigma_{y}=0.75$, $\lambda=2\pi/\omega_0$ [see Eq. (\ref{eq:omega0})]. The
black continuous line corresponds to the electron's trajectory using a Gaussian 
distribution for the charge density of the bunch \cite{wanzenberg2010nonlinear,takayama1982potential},
whereas the green dashed-dotted line corresponds to the approximated model
described in the text [see \eqref{approxerf}]. \label{fig:A-figure-depicting}}
\end{figure}

Let us now consider a model of the electromagnetic field from the dense positron bunch.
The bunches to be used in linear colliders are usually shaped liked sheets, that is they are much
longer than they are wide, and much thinner than they are wide. By assuming
that the bunch propagates along the positive $x$ direction and it lies on the $x$-$z$ plane,
the r.m.s. values of the charge distribution in space are such that $\Sigma_{y}\ll\Sigma_{z}\ll\Sigma_{x}$
(see \figref{A-figure-depicting}). 
We first consider the field in the co-moving frame of the bunch. Here, 
the transverse beam sizes $\Sigma_{y}$ and $\Sigma_{z}$ remain unchanged, 
while the longitudinal becomes longer by a factor of $\gamma_{\text{b}}$, 
which is the Lorentz factor of the bunch, due to the effect of Lorentz 
contraction. Therefore in this co-moving frame the bunches are still sheets. 
The charge density is often modeled as a Gaussian function, that is
\begin{equation}
\rho'(\boldsymbol{r}')=\frac{Ne}{(2\pi)^{3/2}\Sigma_{x}'\Sigma_{y}'\Sigma_{z}'}e^{-\left(\frac{x'^{2}}{2\Sigma_{x}'^{2}}+\frac{y'^{2}}{2\Sigma_{y}'^{2}}+\frac{z'^{2}}{2\Sigma_{z}'^{2}}\right)},\label{eq:chargedensity}
\end{equation}
where primed quantities refer to quantities in the co-moving frame
and $N$ the number of positrons in the bunch. To obtain the electric
field one can use Gauss' law, $\boldsymbol{\nabla}'\cdot\boldsymbol{E}'(\boldsymbol{r}')=4\pi\rho'(\boldsymbol{r}')$,
as explained in \cite{wanzenberg2010nonlinear,takayama1982potential}.
However, since the bunches are flat, there will be a large component of the field
only orthogonal to the sheet, i.e., along the $y'$ (or $y$) direction. Moreover, 
imagining to describe processes occurring at the center of the bunch, we can make 
the leading order expansion of the field component, which results in
\begin{equation}
E_{y'}'(\boldsymbol{r}')\simeq\frac{4\pi Ne}{(2\pi)^{3/2}\Sigma_{x}'\Sigma_{y}'\Sigma_{z}'}y'.\label{eq:approxerf}
\end{equation}
To obtain the field in the lab-frame we perform a Lorentz transformation
with velocity given by $-\boldsymbol{\beta}_{\text{b}}$, with $\boldsymbol{\beta}_{\text{b}}$ 
being the velocity of the bunch. The orthogonal component of the electric field becomes boosted 
by a factor of the bunch Lorentz factor $\gamma_{\text{b}}$, which simplifies 
with the corresponding change in the bunch length $\Sigma_{z}'=\gamma_{\text{b}}\Sigma_{z}$
and one obtains
\begin{equation}
E_{y}(\boldsymbol{r})=\frac{4\pi Ne}{(2\pi)^{3/2}\Sigma_{x}\Sigma_{y}\Sigma_{z}}y.\label{eq:Electriclab}
\end{equation}
A magnetic field arises according to the Lorentz transformation of
the electromagnetic fields: 
\begin{equation}
\boldsymbol{B}_{\perp}(\boldsymbol{r})=-\gamma_{\text{b}}\boldsymbol{\beta}_{\text{b}}\times\boldsymbol{E}'(\boldsymbol{r}')=-\beta_{\text{b}}E_{y}(\boldsymbol{r})\boldsymbol{e}_{3},\label{eq:Bfield}
\end{equation}
where $\boldsymbol{e}_{3}$ is a unit vector in the $z$ direction. 
In order to neglect the dependence of the fields on $x$ and $t$, 
we have implicitly assumed that the electron moves along the $x$
direction at a speed close to the speed of light like the
bunch and that the dynamics in the $y$ direction occurs on a 
timescale much shorter than $\Sigma_{x}$. For the parameters
chosen in \figref{A-figure-depicting}, for example, one can
see that our approximations result in a trajectory very close
to the exact one. Due to the fact that in a number of situations
one can approximate an electromagnetic field as a linear function
of a coordinate, this model is still useful not only as a toy
model for beamstrahlung but also to identify the new regime where 
the transverse motion must be treated quantum mechanically,
instead of classically. In conclusion, the only non-zero
components of the background electromagnetic field in the 
laboratory frame are given by
\begin{align}
E_{y}(\boldsymbol{r})=&\kappa y,\label{eq:Electricapprox}\\
B_{z}(\boldsymbol{r})=&-\beta_{\text{b}}\kappa y,\label{eq:Bfieldapprox}
\end{align}
where
\begin{equation}
\kappa=\frac{2Ne}{\sqrt{2\pi}\Sigma_{x}\Sigma_{y}\Sigma_{z}},\label{eq:kappa}
\end{equation}
is the field gradient.

\section{Classical motion\label{sec:Classical-motion}}

To gain a basic understanding of the problem at hand we first
consider the classical motion in the given field configuration. 
We are interested in the case of an ultrarelativistic electron
with the motion being mainly directed along the
positive $x$-axis, i.e. the $x$ component of the velocity $\boldsymbol{v}$ 
fulfills the condition $v_{x}\simeq 1$, whereas the transverse momentum 
is much smaller than the longitudinal one. In particular, we assume that
the the initial time is set equal to zero and that $v_z(0)=0$, in such a
way that $v_z(t)=0$ for all $t>0$. Since the transverse motion is only
along the $y$ direction, it is convenient to introduce the parameter
\begin{equation}
\xi=\gamma_{0}v_{y,\text{max}},\label{eq:zeta}
\end{equation}
where $\gamma_{0}$ is the initial electron Lorentz gamma factor.
The parameter $\xi$ then becomes on the order of unity when the transverse
motion becomes relativistic. We will restrict ourselves to the (broad)
case where $v_{y,\text{max}}\ll 1$ and therefore $\xi\ll\gamma_{0}$.
The two conditions $v_{x}\simeq 1$ and $\xi\ll\gamma_{0}$ will be employed 
below to solve the equations of motion. We use the Lorentz force equation,
with the electric and magnetic force terms given by
\begin{equation}
q\boldsymbol{E}=\left(\begin{array}{c}
0\\
-e\kappa y\\
0
\end{array}\right),
\end{equation}
and
\begin{align}
\begin{split}
q\boldsymbol{v}\times\boldsymbol{B}&=(-e)\left(\begin{array}{c}
v_{x}\\
v_{y}\\
v_{z}
\end{array}\right)\times\left(\begin{array}{c}
0\\
0\\
1
\end{array}\right)(-\kappa\beta_{\text{b}}y)\\
&=\beta_{\text{b}}e\kappa y\left(\begin{array}{c}
v_{y}\\
-v_{x}\\
0
\end{array}\right),
\end{split}
\end{align}
respectively. The non-vanishing components of the Lorentz equation read
\begin{align}
\frac{dp_{x}}{dt}&=\beta_{\text{b}}e\kappa yv_{y},\\
\frac{dp_{y}}{dt}&=-e\kappa y-e\beta_{\text{b}}\kappa yv_{x}=-(1+\beta_{\text{b}}v_{x})e\kappa y.
\end{align}
Now we use the identity
\begin{equation}
\frac{d\boldsymbol{p}}{dt}=\frac{d\gamma}{dt}m\boldsymbol{v}+\gamma m\frac{d\boldsymbol{v}}{dt},\label{eq:momentumeq}
\end{equation}
where $m$ is the electron mass. By using the equation for the variation of the energy
\begin{equation}
m\frac{d\gamma}{dt}=q\boldsymbol{E}\cdot\boldsymbol{v}=-e\kappa yv_{y},\label{eq:gammaderiv}
\end{equation}
we obtain
\begin{align}
\begin{split}
\gamma m\frac{dv_x}{dt}&=q(\boldsymbol{E}+\boldsymbol{v}\times\boldsymbol{B})_{x}-\frac{d\gamma}{dt}mv_{x}\\
&=\left(\beta_{\text{b}}e\kappa yv_{y}\right)-\left(-e\kappa yv_{y}\right)v_{x}\\
&=(v_{x}+\beta_{\text{b}})e\kappa yv_{y},
\end{split}
\end{align}
and
\begin{align}
\begin{split}
\gamma m\frac{dv_y}{dt}&=q(\boldsymbol{E}+\boldsymbol{v}\times\boldsymbol{B})_{y}-\frac{d\gamma}{dt}mv_{y}\\
&=-e\kappa y-\beta_{\text{b}}e\kappa yv_{x}-(-e\kappa yv_{y})v_{y}\\
&=-e\kappa y\left(1+\beta_{\text{b}}v_{x}-v_{y}^{2}\right).
\end{split}
\end{align}
Now we write
\begin{equation}
v_{x}(t)=1+\delta v_{x}(t),
\end{equation}
and
\begin{equation}
\gamma(t)=\gamma_{0}+\delta\gamma(t),
\end{equation}
where $\gamma_{0}$ is the initial value of the Lorentz gamma factor
for the electron. So we obtain the exact equations of motion as

\begin{align}
\frac{d\delta v_{x}}{dt}&=\frac{1}{\gamma_{0}}\frac{1}{1+\frac{\delta\gamma}{\gamma_{0}}}(\beta_{\text{b}}+1+\delta v_{x})\frac{e\kappa}{m}yv_{y},\\
\frac{dv_{y}}{dt}&=-\frac{1}{\gamma_{0}}\frac{1}{1+\frac{\delta\gamma}{\gamma_{0}}}\left(1+\beta_{\text{b}}+\beta_{\text{b}}\delta v_{x}-v_{y}^{2}\right)\frac{e\kappa}{m}y.
\end{align}
Now we wish to find solutions under the conditions $v_{y}^{2}(t)\ll1$, $\left|\delta v_{x}(t)\right|\ll1$,
$|\delta\gamma(t)|\ll \gamma_{0}$ and $1/\gamma_{\text{b}}^{2}\ll1$.
In this case the equations simplify significantly and we
will verify that the obtained solutions verify these conditions. The
approximate equations of motion then become

\begin{align}
\frac{d\delta v_{x}}{dt}&=(1+\beta_{\text{b}})\frac{e\kappa}{\gamma_{0}m}yv_{y},\\
\frac{dv_{y}}{dt}&=-(1+\beta_{\text{b}})\frac{e\kappa}{\gamma_{0}m}y.
\end{align}
Here, one can replace $\beta_{\text{b}}$ by unity but we prefer to
keep the symbol $\beta_{\text{b}}$ such that we can obtain
the case of a harmonic oscillator via the replacement $\beta_{\text{b}}=0$. 
Now the equation for $y$ can be solved with appropriate initial conditions to obtain
\begin{align}
v_{y}(t)&=\frac{\xi}{\gamma_{0}}\text{cos}(\omega_{0}t),\label{eq:yvelocity}\\
y(t)&=y_{\text{max}}\text{sin}(\omega_{0}t),\label{eq:y-motion}
\end{align}
where
\begin{equation}
\omega_{0}=\sqrt{\frac{(1+\beta_{\text{b}})e\kappa}{\gamma_{0}m}}.\label{eq:omega0}
\end{equation}
And from the definition of Eq. (\ref{eq:zeta}) we obtain that the
amplitude can be expressed in terms of the previously defined quantities
as
\begin{equation}
y_{\text{max}}=\frac{\xi}{\gamma_{0}\omega_{0}}.\label{eq:amplitude}
\end{equation}
Now we can solve the equation for the motion along the $x$ direction.

\begin{align}
\begin{split}
\frac{d\delta v_{x}}{dt}&=\omega_{0}^{2}yv_{y}\\
&=\omega_{0}\left(\frac{\xi}{\gamma_{0}}\right)^{2}\text{sin}(\omega_{0}t)\text{cos}(\omega_{0}t)\\
&=\frac{\omega_{0}}{2}\left(\frac{\xi}{\gamma_{0}}\right)^{2}\text{sin}(2\omega_{0}t),\label{eq:x-acc}
\end{split}
\end{align}
and upon integration we obtain
\begin{equation}
\delta v_{x}(t)=-\frac{1}{4}\left(\frac{\xi}{\gamma_{0}}\right)^{2}\text{cos}(2\omega_{0}t)+C_{1},\label{eq:x-vel}
\end{equation}
where $C_{1}$ is a constant of integration. Now, the constant difference
between $\delta v_{x}(t)$ and $v_{x}(t)$ can be absorbed in the
constant of integration $C_{1}$ in Eq. (\ref{eq:x-vel}) such that
\begin{equation}
v_{x}(t)=-\frac{1}{4}\left(\frac{\xi}{\gamma_{0}}\right)^{2}\text{cos}(2\omega_{0}t)+C_{2}.
\end{equation}
In order to determine the constant $C_{2}$, we can use $v_{x}^{2}(0)+v_{y}^{2}(0)=v_{0}^{2}=1-1/\gamma_{0}^{2}$. Thus, $v_{x}^{2}(0)=1-(1+\xi^{2})/\gamma_{0}^{2}$ and $v_{x}(0)\simeq1-(1+\xi^{2})/2\gamma_{0}^{2}$, we can determine the constant of integration $C_{2}$ and we obtain:
\begin{equation}
v_{x}(t)=1-\frac{1}{2\gamma_{0}^{2}}-\frac{\xi^{2}}{4\gamma_{0}^{2}}-\frac{\xi^{2}}{4\gamma_{0}^{2}}\text{cos}(2\omega_{0}t),\label{eq:x-vel-1}
\end{equation}
which upon integration yields
\begin{equation}
x(t)=\left(1-\frac{1}{2\gamma_{0}^{2}}-\frac{\xi^{2}}{4\gamma_{0}^{2}}\right)t-\frac{\xi^{2}}{4\gamma_{0}^{2}}\frac{\text{sin}(2\omega_{0}t)}{2\omega_{0}}.\label{eq:x(t)}
\end{equation}
Now, it is clear that the conditions for the approximate solutions
are fulfilled as long as $\xi\ll\gamma_{0}$. However we must use
Eq. (\ref{eq:gammaderiv}) to check when the condition $|\delta\gamma(t)|\ll\gamma_{0}$
is fulfilled. Using Eq. (\ref{eq:gammaderiv}) we obtain

\begin{equation}
\frac{d\gamma}{dt}=-\frac{\omega_{0}}{1+\beta_{\text{b}}}\frac{\xi^{2}}{2\gamma_{0}}\text{sin}(2\omega_{0}t).
\end{equation}
And so we can integrate to obtain
\begin{equation}
\gamma(t)=\gamma_{0}+\frac{1}{1+\beta_{\text{b}}}\frac{\xi^{2}}{4\gamma_{0}}\left[\text{cos}(2\omega_{0}t)-1\right],
\end{equation}
and therefore 
\begin{equation}
\frac{\delta\gamma(t)}{\gamma_{0}}=\frac{1}{1+\beta_{\text{b}}}\frac{\xi^{2}}{4\gamma_{0}^{2}}\left[\text{cos}(2\omega_{0}t)-1\right].
\end{equation}
In this way the condition $|\delta\gamma(t)|\ll\gamma_{0}$ is equivalent to 
the condition $\xi\ll\gamma_{0}$. Therefore, as long as this condition
is fulfilled the neglected terms are smaller than a factor of at least
$\xi/\gamma_{0}$ compared to the dominant ones.

\section{Solution of Dirac equation\label{sec:Solution-of-Dirac-1}}

Classically the fields of Eq. (\ref{eq:Electricapprox}) and (\ref{eq:Bfieldapprox})
gives a force as in Hooke's law and therefore harmonic oscillations
as seen in section (\ref{sec:Classical-motion}). Harmonic oscillator
wave functions should therefore be involved in the solution of the
Dirac equation. We have $\boldsymbol{E}(\boldsymbol{r})=-\boldsymbol{\nabla}\varphi(\boldsymbol{r})$
so due to the simple structure of the electric field in Eq. (\ref{eq:Electricapprox})
the potential depends only on the y-coordinate, i.e. we have

\begin{equation}
\varphi (y)=-\frac{\kappa y^{2}}{2}.
\end{equation}
A vector potential which gives us the magnetic field of Eq. (\ref{eq:Bfieldapprox}) is

\begin{equation}
\boldsymbol{A}(y)=\frac{\beta_{\text{b}}\kappa}{2}(y^{2},0,0).
\end{equation}
The Dirac equation in an external field reads

\begin{equation}
\left(\hat{\slashed p}+e\slashed A(\boldsymbol{r})-m\right)\psi(\boldsymbol{r},t)=0,\label{eq:Diraceq}
\end{equation}
where $\psi(\boldsymbol{r},t)$ is the electron bispinor wave function.
This can also be rewritten as

\begin{equation}
i\frac{\partial\psi(\boldsymbol{r},t)}{\partial t}=\hat{H}\psi(\boldsymbol{r},t),\label{eq:Diraceq-1}
\end{equation}
with

\begin{equation}
\hat{H}=\boldsymbol{\alpha}\cdot\hat{\boldsymbol{\Pi}}-e\varphi (\boldsymbol{r})+\gamma^{0}m,
\end{equation}
where $\boldsymbol{\alpha}^{i}=\gamma^{0}\gamma^{i}$,
$i=1,2,3$, and $\boldsymbol{\Pi}_{i}=\boldsymbol{\hat{p}}_{i}+e\boldsymbol{A}_{i}(\boldsymbol{r})$
(for electron). We consider a problem where the potentials have no time dependence
and then take the usual approach \cite{LLQED} of finding the stationary
states and write

\begin{equation}
\psi(\boldsymbol{r},t)=e^{-i\varepsilon t}\left(\begin{array}{c}
\phi(\boldsymbol{r})\\
\chi(\boldsymbol{r})
\end{array}\right),
\end{equation}
where $\varepsilon$ will be the energy. This leads to

\begin{equation}
(\varepsilon-V(\boldsymbol{r})-m)\phi(\boldsymbol{r})=\boldsymbol{\sigma}\cdot(-i\boldsymbol{\nabla}+e\boldsymbol{A}(\boldsymbol{r}))\chi(\boldsymbol{r}),\label{eq:eqphi}
\end{equation}

\begin{equation}
(\varepsilon-V(\boldsymbol{r})+m)\chi(\boldsymbol{r})=\boldsymbol{\sigma}\cdot(-i\boldsymbol{\nabla}+e\boldsymbol{A}(\boldsymbol{r}))\phi(\boldsymbol{r}),\label{eq:eqchi}
\end{equation}
where $\boldsymbol{\sigma}$ denote the three Pauli matrices. 
Now from Eq. (\ref{eq:eqphi}) we find

\begin{equation}
\chi(\boldsymbol{r})=\frac{1}{\varepsilon+e\varphi (y)+m}\boldsymbol{\sigma}\cdot\hat{\boldsymbol{\Pi}}\phi(\boldsymbol{r}),
\end{equation}
and inserting this in Eq. (\ref{eq:eqchi}) we obtain a differential
equation for $\phi(\boldsymbol{r})$

\begin{align}
\begin{split}
&(\varepsilon+e\varphi (y)-m)\phi(\boldsymbol{r})\\
&=\boldsymbol{\sigma}\cdot(-i\boldsymbol{\nabla}+e\boldsymbol{A}(y))\frac{1}{\varepsilon+e\varphi (y)+m}\boldsymbol{\sigma}\cdot\hat{\boldsymbol{\Pi}}\phi(\boldsymbol{r}).\label{eq:phiequnreduced}
\end{split}
\end{align}
To find the solution for $\phi(\boldsymbol{r})$ we need to rewrite this such that we can isolate the Laplacian of $\phi(\boldsymbol{r})$. The product rule for the gradient gives
us a term where it acts on $(\varepsilon+e\varphi (y)+m)^{-1}$ and one
where it acts on $\boldsymbol{\sigma}\cdot\hat{\boldsymbol{\Pi}}\phi(\boldsymbol{r})$
so this gives us

\begin{align}
\begin{split}
&(\varepsilon+e\varphi (y)-m)\phi(\boldsymbol{r})\\
&=-i\boldsymbol{\sigma}\cdot\boldsymbol{\nabla}\left(\frac{1}{\varepsilon+e\varphi (y)+m}\right)\boldsymbol{\sigma}\cdot\hat{\boldsymbol{\Pi}}\phi(\boldsymbol{r})\\
&-\frac{1}{\varepsilon+e\varphi (y)+m}i\boldsymbol{\sigma}\cdot\boldsymbol{\nabla}\left(\boldsymbol{\sigma}\cdot\hat{\boldsymbol{\Pi}}\phi(\boldsymbol{r})\right)\\
&+\boldsymbol{\sigma}\cdot e\boldsymbol{A}(y)\frac{1}{\varepsilon+e\varphi (y)+m}\boldsymbol{\sigma}\cdot\hat{\boldsymbol{\Pi}}\phi(\boldsymbol{r})\\
&=-i\boldsymbol{\sigma}\cdot\boldsymbol{\nabla}\left(\frac{1}{\varepsilon+e\varphi (y)+m}\right)\boldsymbol{\sigma}\cdot\hat{\boldsymbol{\Pi}}\phi(\boldsymbol{r})\\
&+\frac{1}{\varepsilon+e\varphi (y)+m}\boldsymbol{\sigma}\cdot\left(-i\boldsymbol{\nabla}+e\boldsymbol{A}(y)\right)\boldsymbol{\sigma}\cdot\hat{\boldsymbol{\Pi}}\phi(\boldsymbol{r})\\
&=-i\boldsymbol{\sigma}\cdot\boldsymbol{\nabla}\left(\frac{1}{\varepsilon+e\varphi (y)+m}\right)\boldsymbol{\sigma}\cdot\hat{\boldsymbol{\Pi}}\phi(\boldsymbol{r})\\
&+\frac{1}{\varepsilon+e\varphi (y)+m}\left[\boldsymbol{\sigma}\cdot\hat{\boldsymbol{\Pi}}\right]^{2}\phi(\boldsymbol{r})\\
&=i\boldsymbol{\sigma}_{y}\cdot\frac{e\varphi'(y)}{\left(\varepsilon+e\varphi (y)+m\right)^{2}}\boldsymbol{\sigma}\cdot\hat{\boldsymbol{\Pi}}\phi(\boldsymbol{r})\\
&+\frac{1}{\varepsilon+e\varphi (y)+m}\left[\boldsymbol{\sigma}\cdot\hat{\boldsymbol{\Pi}}\right]^{2}\phi(\boldsymbol{r}).
\end{split}
\end{align}
Multiplying by $\left(\varepsilon+e\varphi (y)+m\right)$ on both sides
we obtain

\begin{align}
\begin{split}
&\left(\left(\varepsilon+e\varphi (y)\right)^{2}-m^{2}\right)\phi(\boldsymbol{r})\\
&=i\boldsymbol{\sigma}_{y}\cdot\frac{1}{\varepsilon+e\varphi (y)+m}e \varphi'(y) \boldsymbol{\sigma}\cdot\hat{\boldsymbol{\Pi}}\phi(\boldsymbol{r})+\left[\boldsymbol{\sigma}\cdot\hat{\boldsymbol{\Pi}}\right]^{2}\phi(\boldsymbol{r})\\
&=-i\frac{1}{\varepsilon+e\varphi (y)+m}\left(\boldsymbol{\sigma}_{y}\cdot eE_{y}(y)\right)\boldsymbol{\sigma}\cdot\hat{\boldsymbol{\Pi}}\phi(\boldsymbol{r})+\left[\boldsymbol{\sigma}\cdot\hat{\boldsymbol{\Pi}}\right]^{2}\phi(\boldsymbol{r})\\
&=\frac{-i}{\varepsilon+e\varphi (y)+m}\left(eE_{y}(y)\hat{\boldsymbol{\Pi}}_{y}+i\boldsymbol{\sigma}\cdot(e\boldsymbol{E}(y)\times\hat{\boldsymbol{\Pi}})\right)\phi(\boldsymbol{r})\\
&+\left[\boldsymbol{\sigma}\cdot\hat{\boldsymbol{\Pi}}\right]^{2}\phi(\boldsymbol{r}).
\end{split}
\end{align}
Now we need to consider the term $\left[\boldsymbol{\sigma}\cdot\hat{\boldsymbol{\Pi}}\right]^{2}$ by letting it act on a test function
 $f$.

\begin{align}
\begin{split}
&\left[\boldsymbol{\sigma}\cdot(\boldsymbol{\hat{p}}+e\boldsymbol{A}(y))\right]\left[\boldsymbol{\sigma}\cdot(\boldsymbol{\hat{p}}f+e\boldsymbol{A}(y)f)\right]\\
&=\boldsymbol{\hat{p}}^{2}f+\boldsymbol{\sigma}\cdot\left(\boldsymbol{\nabla}\times\left[e\boldsymbol{A}(y)\right]\right)f\\
&+2eA_{x}(y)\hat{p}_{x}f+e^{2}A^{2}(y)f.
\end{split}
\end{align}
Then, finally, we obtain

\begin{align}
\begin{split}
&\left[\boldsymbol{\hat{p}}^{2}+e\boldsymbol{\sigma}\cdot\boldsymbol{B}(y)+2eA_{x}(y)\hat{p}_{x}\right.\\
&-2\varepsilon e\varphi (y)+e^{2}A^{2}(y)-e^{2}\varphi^{2}(y)\\
&-i\frac{1}{\varepsilon+e\varphi (y)+m}\left(eE_{y}\hat{\Pi}_{y}+i\boldsymbol{\sigma}\cdot(e\boldsymbol{E}(y)\times\hat{\boldsymbol{\Pi}})\right)\\
&\left.-\left(\varepsilon^{2}-m^{2}\right)\right]\phi(\boldsymbol{r})=0.\label{eq:exactphi}
\end{split}
\end{align}
This is the exact differential equation for $\phi(\boldsymbol{r})$.
In Eq. (\ref{eq:exactphi}) the dominant terms driving the dynamics
are the terms $2eA_{x}(y)\hat{p}_{x}-2\varepsilon e\varphi (y)$. We will
try with a separable solution which yields free motion in the $x$-direction
such that this becomes $2eA_{x}(y)p_{x}$ (no longer an operator) which
is under most circumstances nearly equal the other term $-2\varepsilon e\varphi (y)$.
These two terms correspond to the adding of the electric and magnetic
forces on the particle. It turns out that in the regime we are interested
in, many of the terms in Eq. (\ref{eq:exactphi}) can be neglected. 

\subsection{$e^2\varphi^{2}(y)$ term\label{subsec:-termA}}

If $e^2\varphi^{2}(y)$ should be much smaller than $-2\varepsilon e\varphi (y)$ we
should have that $2\varepsilon\gg -e\varphi (y)$. The most stringent condition
is then to use $-e\varphi_{\text{max}}=-e\varphi(y_{\text{max}})=e\kappa y_{\text{max}}^{2}/2=e\kappa\xi^{2}/2\gamma_{0}^{2}\omega_{0}^{2}$.
Our condition then becomes $1\gg\frac{\xi^{2}}{\gamma_{0}^{2}}\frac{1}{4(1+\beta_{\text{b}})}$,
a condition which will be fulfilled as we require exactly that $\xi\ll\gamma_{0}$.
The same argument goes for the $e^2A^2(y)$ term.

\subsection{$e\boldsymbol{\sigma}\cdot\boldsymbol{B}(y)$ term\label{subsec:-termB}}

Here we should have $-2\varepsilon e\varphi (y)\gg eB_{z}(y)$
corresponding to

\begin{equation}
2\varepsilon\frac{e\kappa y^{2}}{2}\gg e\kappa y,
\end{equation}
which reduces to

\begin{equation}
y\gg\frac{1}{\varepsilon},\label{eq:termBcond}
\end{equation}
which is seen to be roughly the Compton wavelength divided by a factor of $\gamma_{0}$. The problem could, in fact, be solved while including this term, and the effect would be that the spin up and spin down wave-functions are shifted by the distance $1/\varepsilon$ compared to eachother. This is however completely negligible. As we will see later, the transition to the new regime happens when the typical length of the problem becomes on the order of the Compton wavelength, and this condition is a factor of $\gamma_{0}$ below this.

\subsection{$\frac{eE_{y}\hat{\Pi}_{y}}{\varepsilon+e\varphi (y)+m}$ term}

We obtain the most stringent condition by inserting the maximum value of the classical momentum and so
\begin{align}
\frac{eE_{y}\hat{\Pi}_{y}}{\varepsilon+e\varphi (y)+m}\simeq\frac{e\kappa yp_{y,\text{max}}}{\varepsilon}\simeq\frac{e\kappa ym\xi}{\varepsilon}.
\end{align}
We should then have

\begin{equation}
\frac{e\kappa ym\xi}{\varepsilon}\ll-2\varepsilon e\varphi (y)=\varepsilon e\kappa y^{2},
\end{equation}
which reduces to

\begin{equation}
y\gg \frac{\xi}{\gamma_{0}}\frac{1}{\varepsilon},\label{eq:condC}
\end{equation}
and since we require that $\gamma_{0}\gg\xi$ if Eq. (\ref{eq:termBcond})
is fulfilled, then so is Eq. (\ref{eq:condC}).

\subsection{$\frac{\boldsymbol{\sigma}\cdot(e\boldsymbol{E}\times\hat{\boldsymbol{\Pi}})}{\varepsilon+e\varphi (y)+m}$}

We have that $\boldsymbol{\sigma}\cdot(\boldsymbol{E}\times\hat{\boldsymbol{\Pi}}) = \boldsymbol{\sigma}_{x} E_{y}(y)p_{z}-\boldsymbol{\sigma}_{z} E_{y}(y)(p_{x}+eA_{x}(y))$
. The previous terms have either had the matrix structure of the identity or $\boldsymbol{\sigma}_{z}$ while here we also have a term proportional to $\boldsymbol{\sigma}_{x}$ i.e. a mixing between the spin states. The $\boldsymbol{\sigma}_{z}$ term is on the same size as the one
from  \subsecref{-termB} and therefore negligible. The mixing term will be even smaller as $p_z$ will be 0 initially and on the order of $m$ in the final state, so a factor of $\gamma_{0}$ smaller than the already negligible small correction. 

Now that we have argued for the smallness of the additional terms
we are left with the equation

\begin{equation}
\left[\boldsymbol{\hat{p}}^{2}+2eA_{x}(y)\hat{p}_{x}-2\varepsilon e\varphi (y)-\left(\varepsilon^{2}-m^{2}\right)\right]\phi(\boldsymbol{r})=0.\label{eq:phidiff}
\end{equation}
This gives us

\begin{equation}
\left[-\boldsymbol{\nabla}^{2}+e\kappa\beta y^{2}(-i\partial_{x})+\varepsilon e\kappa y^{2}-(\varepsilon^{2}-m^{2})\right]\phi(\boldsymbol{r})=0.
\end{equation}
To solve this equation we try the ansatz $\phi(\boldsymbol{r})=I(y)e^{ip_{x}x+ip_{z}z}\boldsymbol{s}$
where $\boldsymbol{s}$ is any 2-component vector and so we obtain the following differential equation only in the $y$-coordinate,

\begin{equation}
\left[-\frac{d^{2}}{d y^{2}}+e\kappa(\beta_{\text{b}} p_{x}+\varepsilon)y^{2}-(\varepsilon^{2}-p_{x}^{2}-p_{z}^{2}-m^{2})\right]I(y)=0.
\end{equation}
By defining

\begin{equation}
\frac{1}{L}=\sqrt[4]{e\kappa(\beta_{\text{b}} p_{x}+\varepsilon)},
\end{equation}
and introducing the dimensionless variable,
\begin{equation}
\eta=y/L,
\end{equation}
we obtain that

\begin{equation}
\left[\frac{d^{2}}{d\eta^{2}}-\eta^{2}+L^{2}\left(\varepsilon^{2}-p_{x}^{2}-p_{z}^{2}-m^{2}\right)\right]I(\eta)=0,\label{eq:eq51}
\end{equation}
and defining

\begin{equation}
a=L^{2}\left(\varepsilon^{2}-p_{x}^{2}-p_{z}^{2}-m^{2}\right),
\end{equation}
\eqref{eq51} becomes

\begin{equation}
\left[\frac{d^{2}}{d\eta^{2}}-\eta^{2}+a\right]I(\eta)=0,
\end{equation}
which has normalizable solutions when $a=2n+1$ with $n$ integer (see e.g. \cite{bagrov1990exact,griffiths2016}), which
we denote $I_{n}(\eta)$. And so the solutions to Eq. (\ref{eq:phidiff})
are given by

\begin{equation}
\phi_{n}(\boldsymbol{r})=e^{i(p_{x}x+p_{z}z)}I_{n}(\eta(y))\boldsymbol{s},
\end{equation}
when the constants are related by

\begin{equation}
\varepsilon_{n}^{2}=\frac{1}{L^{2}}\left(2n+1\right)+p_{x}^{2}+p_{z}^{2}+m^{2},\label{eq:energyrelation}
\end{equation}
and the solutions can be written explicitly as

\begin{equation}
I_{n}(\eta)=N_{n}e^{-\eta^{2}/2}H_{n}(\eta),
\end{equation}
where $N_{n}$ is a normalization constant to be found and $H_{n}(\eta)$
are the Hermite polynomials normalized such that $\int_{-\infty}^{\infty}H_{m}(x)H_{n}(x)e^{-x^{2}}dx=\sqrt{\pi}2^{n}n!\delta_{nm}$ where $\delta_{nm}$ is the Kronecker delta function. We would like to normalize $I_{n}(\eta)$ such that 
\begin{equation}
\int|I_{n}(\eta)|^{2}dy=1,
\end{equation}
which gives us that

\begin{equation}
N_{n}=\frac{1}{\sqrt{2^{n}L\sqrt{\pi}n!}}.
\end{equation}
So we have our solutions to the Dirac equation as

\begin{equation}
\psi(\boldsymbol{r},t)=\frac{1}{\sqrt{2L_{x}L_{z}}}\left(\begin{array}{c}
I_{n}(\eta)\boldsymbol{s}\\
\frac{\boldsymbol{\sigma}\cdot(-i\boldsymbol{\nabla}+e\boldsymbol{A}(y))}{\varepsilon+e\varphi (y)+m}I_{n}(\eta)\boldsymbol{s}
\end{array}\right)e^{i(p_{x}x+p_{z}z-\varepsilon_{n}t)},\label{eq:firstsol}
\end{equation}
where $L_{x}L_{z}$ is a normalization area in the $xz$ plane (1
particle per area). To obtain a more explicit expression for the lower
two components in the bispinor of Eq. (\ref{eq:firstsol}) we insert
the Pauli matrices to obtain

\begin{align}
\begin{split}
&\boldsymbol{\sigma}\cdot(-i\boldsymbol{e}_2\frac{d}{dy}+e\boldsymbol{A}(y))I_{n}(\eta)\\
&=y^{2}I_{n}(\eta)\frac{e\kappa\beta_\text{b}}{2}\boldsymbol{\sigma}_{x}-i\boldsymbol{\sigma}_{y}\frac{dI_{n}(\eta)}{dy}\\
&=\left(\begin{array}{cc}
0 & I_{n}(\eta)\eta^{2}C-\frac{1}{L}\frac{dI_{n}}{d\eta}\\
I_{n}(\eta)\eta^{2}C+\frac{1}{L}\frac{dI_{n}}{d\eta} & 0
\end{array}\right)
\end{split}
\end{align}
where $\boldsymbol{e}_2$ is a unit vector in the $y$-direction and we defined $C=\frac{e\kappa\beta_\text{b} L^{2}}{2}$. Calculating the
spin dependent part of Eq. (\ref{eq:firstsol}), we define

\begin{equation}
U_{\uparrow}(y)=\left(\begin{array}{c}
I_{n}(\eta)\\
0\\
\frac{p_{z}I_{n}(\eta)}{\varepsilon_{n}+e\varphi (y)+m}\\
\frac{p_{x}I_{n}(\eta)+I_{n}(\eta)\eta^{2}C+\frac{1}{L}\frac{dI_{n}(\eta)}{d\eta}}{\varepsilon_{n}+e\varphi (y))+m}
\end{array}\right),
\end{equation}

\begin{equation}
U_{\downarrow}(y)=\left(\begin{array}{c}
0\\
I_{n}(\eta)\\
\frac{p_{x}I_{n}(\eta)+I_{n}(\eta)\eta^{2}C-\frac{1}{L}\frac{dI_{n}(\eta)}{d\eta}}{\varepsilon_{n}+e\varphi (y)+m}\\
\frac{-p_{z}I_{n}(\eta)}{\varepsilon_{n}+e\varphi (y)+m}
\end{array}\right).
\end{equation}
Finally, since the potential $(-e\varphi (y))$ in the denominator is much smaller
than the energy $\varepsilon_{n}$ (this is the same condition as in subsection \ref{subsec:-termB}) we can make the following approximation,
valid when $\gamma_{0}\gg\xi$

\begin{align}
\begin{split}
&\frac{p_{x}I_{n}(\eta)+I_{n}\eta^{2}C+\frac{1}{L}\frac{dI_{n}(\eta)}{d\eta}}{\varepsilon_{n}+e\varphi (y)+m}\\
&=\frac{p_{x}I_{n}(\eta)+I_{n}(\eta)\eta^{2}C+\frac{1}{L}\frac{dI_{n}(\eta)}{d\eta}}{\left(\varepsilon_{n}+m\right)(1+\frac{e\varphi (y)}{\varepsilon_{n}+m})}\\
&\simeq\frac{p_{x}I_{n}(\eta)+I_{n}(\eta)\eta^{2}C+\frac{1}{L}\frac{dI_{n}(\eta)}{d\eta}}{\left(\varepsilon_{n}+m\right)}(1-\frac{e\varphi (y)}{\varepsilon_{n}+m})\\
&\simeq\frac{p_{x}I_{n}(\eta)-\frac{p_{x}}{\varepsilon_{n}+m}e\varphi (y) I_{n}(\eta)+I_{n}(\eta)\eta^{2}C+\frac{1}{L}\frac{dI_{n}(\eta)}{d\eta}}{\varepsilon_{n}+m}\\
&\simeq\frac{p_{x}I_{n}(\eta)+I_{n}(\eta)\eta^{2}D+\frac{1}{L}\frac{dI_{n}(\eta)}{d\eta}}{\varepsilon_{n}+m},
\end{split}
\end{align}
where $D=\frac{e\kappa(\frac{p_{x}}{\varepsilon_{n}+m}+\beta_{\text{b}})L^{2}}{2}\simeq\frac{e\kappa(1+\beta_{\text{b}})L^{2}}{2}$,
and so we obtain 

\begin{equation}
U_{\uparrow}(y)=\left(\begin{array}{c}
I_{n}(\eta)\\
0\\
\frac{p_{z}I_{n}(\eta)}{\varepsilon_{n}+m}\\
\frac{p_{x}I_{n}(\eta)+I_{n}(\eta)\eta^{2}D+\frac{1}{L}\frac{dI_{n}(\eta)}{d\eta}}{\varepsilon_{n}+m}
\end{array}\right),
\end{equation}

\begin{equation}
U_{\downarrow}(y)=\left(\begin{array}{c}
0\\
I_{n}(\eta)\\
\frac{p_{x}I_{n}(\eta)+I_{n}(\eta)\eta^{2}D-\frac{1}{L}\frac{dI_{n}(\eta)}{d\eta}}{\varepsilon_{n}+m}\\
\frac{-p_{z}I_{n}(\eta)}{\varepsilon_{n}+m}
\end{array}\right),
\end{equation}
such that our solutions, finally, can be spanned by the two solutions

\begin{equation}
\psi_{\uparrow}(\boldsymbol{r},t)=\frac{1}{\sqrt{2L_{x}L_{z}}}e^{i(p_{x}x+p_{z}z-\varepsilon_{n}t)}U_{\uparrow}(y)\label{eq:spinup}
\end{equation}

\begin{equation}
\psi_{\downarrow}(\boldsymbol{r},t)=\frac{1}{\sqrt{2L_{x}L_{z}}}e^{i(p_{x}x+p_{z}z-\varepsilon_{n}t)}U_{\downarrow}(y)\label{eq:spindown}
\end{equation}
It was earlier stated that the normalization was for one particle
per area, however unproven. To show that this is correct within our
approximation, we calculate $\int\psi^{\dagger}(\boldsymbol{r},t)\psi(\boldsymbol{r},t) dV$. We have

\begin{align}
\begin{split}
&\int\psi_{\uparrow}^{\dagger}(\boldsymbol{r},t)\psi_{\uparrow}(\boldsymbol{r},t)dV\\
&=\frac{1}{2}\int dy\left(|I_{n}(\eta)|^{2}+\left[\frac{p_{x}I_{n}(\eta)+I_{n}(\eta)\eta^{2}C+\frac{1}{L}\frac{dI_{n}(\eta)}{d\eta}}{\varepsilon_{n}+e\varphi (y)+m}\right]^{2}\right),
\end{split}
\end{align}
where the integration over $dxdz$ has canceled out with the area
$L_{x}L_{z}$ in the front factor of Eq. (\ref{eq:spinup}). We show
here the case up spin-up, but the result is the same for spin-down.

By using the properties that the derivative of the $I_{n}(\eta)$
function, as they are harmonic oscillator wave functions, is related
to the stepped up and down wave-functions we find that we should add
a normalization factor on the wave function of

\begin{equation}
\frac{1}{\sqrt{1+\frac{p_{x}^{2}-\left(E_{n}+m\right)^{2}+\left(\frac{3}{2}n^{2}+\frac{3}{2}n+\frac{3}{4}\right)D^{2}+\left(\frac{1}{L^{2}}+p_{x}D\right)\left(n+\frac{1}{2}\right)}{2\left(\varepsilon_{n}+m\right)^{2}}}}.
\end{equation}
To estimate the size of this correction we also need to know the typical size of $n$. The solutions 
$I_{n}(\eta)$ rapidly drop off, for large $n$, when beyond the classical turning points which can be found from Eq. (\ref{eq:energyrelation}) and (\ref{eq:phidiff}) to correspond to $\eta^2 > 2n+1$.
We have that $e(A_{x}(y)-\varphi(y)) \sim D\eta^2$ and therefore $nD \sim -e\varphi(y_{\text{max}})$. Thus we have terms that are on the order of $-e\varphi(y_{\text{max}})/\varepsilon_n$ or this factor squared, which is therefore of the same size as the correction found in section \ref{subsec:-termA}.  The term $n/L^2$ can be seen to be on the same order by replacing $n \sim -e\varphi(y_{\text{max}})/D$ and using the definition of $L$. Therefore, as long as $\gamma_{0}\gg\xi$ the normalization of Eq. (\ref{eq:spinup})
and (\ref{eq:spindown}) is correct within our accuracy.

\section{Radiation emission\label{sec:rademit}}

Now that we have obtained the wave-functions we can calculate the
probability of radiation emission by using the transition matrix element
from an initial state $\psi_{i}(x)$ to a final state $\psi_{f}(x)$
while emitting a photon with momentum four-vector $k^{\mu}=(\omega,\boldsymbol{k})$ and polarization
$\epsilon$, which is given by

\begin{equation}
S_{fi}=\int d^{4}x\bar{\psi}_{f}(x)ie\sqrt{\frac{4\pi}{2\omega V}}\slashed{\epsilon}^{*}e^{ikx}\psi_{i}(x)\label{eq:matrixelement}.
\end{equation}
Then the differential rate of emission $dW$ is usually given by

\begin{equation}
dW=\left|S_{fi}\right|^{2}\frac{1}{T}\frac{Vd^{3}p_{f}}{(2\pi)^{3}}\frac{Vd^{3}k}{(2\pi)^{3}},\label{eq:propability}
\end{equation}
where $V$ is the normalization volume and $T$ the interaction time, factors which eventually cancel out.
In our case the density of final states of the electron instead becomes
$\frac{Vd^{3}p_{f}}{(2\pi)^{3}}\rightarrow\frac{dp_{x}dp_{z}L_{x}L_{z}}{(2\pi)^{2}}\sum_{n_{f}}$,
where $L_{x}L_{z}$ is a normalization area. This change is due simply
to the fact that one quantum number is discrete instead of continuous.
Inserting our wave-functions from Eq. (\ref{eq:spindown}) we obtain

\begin{align}
\begin{split}
S_{fi}&=ie\sqrt{\frac{4\pi}{2\omega V}}\frac{1}{2L_{x}L_{z}}\int d^{4}x\bar{U}_{f}(y)\slashed{\epsilon}^{*}U_{i}(y)\\
&\times e^{-ik_{y}y}e^{i(p_{x,i}-p_{x,f}-k_{x})x}e^{i(p_{z,i}-p_{z,f}-k_{z})z}\\
&\times e^{i(\varepsilon_{f}+\omega-\varepsilon_{i})t},
\end{split}
\end{align}
and carrying out the trivial integrations we obtain,

\begin{align}
\begin{split}
S_{fi}&=ie\sqrt{\frac{4\pi}{2\omega V}}\frac{1}{2L_{x}L_{z}}(2\pi)^{3}\int\bar{U}_{f}(y)\slashed{\epsilon}^{*}U_{i}(y)e^{-ik_{y}y}dy\\
&\times\delta(p_{x,i}-p_{x,f}-k_{x})\delta(p_{z,i}-p_{z,f}-k_{z})\\
&\times\delta(\varepsilon_{f}+\omega-\varepsilon_{i}).
\end{split}
\end{align}
Since we need this quantity squared, we must consider the meaning
of the delta-function squared. Here we take the usual approach to
obtain factors of the normalization volume and time, i.e. $\left[\delta(p_{x,i}-p_{x,f}-k_{x})\delta(p_{z,i}-p_{z,f}-k_{z})\delta(\varepsilon_{f}+\omega-\varepsilon_{i})\right]^{2}=\frac{L_{x}L_{z}T}{(2\pi)^{3}}\delta(p_{x,i}-p_{x,f}-k_{x})\delta(p_{z,i}-p_{z,f}-k_{z})\delta(\varepsilon_{f}+\omega-\varepsilon_{i})$
and so we obtain

\begin{align}
\begin{split}
|S_{fi}|^{2}&=\frac{4\pi e^{2}}{2\omega V}\frac{1}{(2L_{x}L_{z})^{2}}(2\pi)^{6}\\
&\times\left|\int\bar{U}_{f}(y)\slashed{\epsilon}^{*}U_{i}(y)e^{-ik_{y}y}dy\right|^{2}\\
&\times\frac{L_{x}L_{z}T}{(2\pi)^{3}}\delta(p_{x,i}-p_{x,f}-k_{x})\delta(p_{z,i}-p_{z,f}-k_{z})\\
&\times\delta(\varepsilon_{f}+\omega-\varepsilon_{i}).
\end{split}
\end{align}
Now integrating over final electron momentum we obtain

\begin{align}
\begin{split}
&\int\frac{dp_{x}dp_{z}L_{x}L_{z}}{(2\pi)^{2}}\sum_{n_{f}}|S_{fi}|^{2}\\
&=\sum_{n_{f}}\frac{4\pi e^{2}}{2\omega V}\frac{1}{(2L_{x}L_{z})^{2}}(2\pi)^{6}\\
&\times\left|\int\bar{U}_{f}(y)\slashed{\epsilon}^{*}U_{i}(y)e^{-ik_{y}y}dy\right|^{2}\\
&\times\frac{L_{x}L_{z}T}{(2\pi)^{3}}\frac{L_{x}L_{z}}{(2\pi)^{2}}\delta(\varepsilon_{f}+\omega-\varepsilon_{i})\\
&=\sum_{n_{f}}\frac{e^{2}}{4\omega V}(2\pi)^{2}\left|\int\bar{U}_{f}(y)\slashed{\epsilon}^{*}U_{i}(y)e^{-ik_{y}y}dy\right|^{2}\\
&\times T\delta(\varepsilon_{f}+\omega-\varepsilon_{i}).
\end{split}
\end{align}
Now we must only add the photon density of states from Eq. (\ref{eq:propability})
and we obtain the differential rate as

\begin{align}
\begin{split}
dW&=\sum_{n_{f}}\frac{e^{2}}{8\pi\omega}\left|\int\bar{U}_{f}(y)\slashed{\varepsilon}^{*}U_{i}(y)e^{-ik_{y}y}dy\right|^{2}\\
&\times\delta(\varepsilon_{f}+\omega-\varepsilon_{i})\omega{}^{2}d\omega d\Omega.
\end{split}
\end{align}
Now we wish to integrate over $d\text{cos}\theta$ in $d\Omega=d\Phi d\text{cos}\theta$
to get rid of the last delta-function. To do this we use the energy
relation of Eq. (\ref{eq:energyrelation}) to write the final energy
and use that the momentum delta functions have fixed $p_{x,f}=p_{x,i}-k_{x}$
and $p_{z,f}=p_{z,i}-k_{z}$. Now writing the photon momentum vector
$\boldsymbol{k}$ in spherical coordinates,

\begin{equation}
\boldsymbol{k}=\omega(\text{cos}\theta,\text{sin}\theta\text{cos}\Phi,\text{sin}\theta\text{sin}\Phi),
\end{equation}
we have that
\begin{align}
p_{z,f}=&-k_{z}=-\omega\text{sin}\theta\text{sin}\Phi,\\
p_{x,f}=&p_{x,i}-\omega\text{cos}\theta.
\end{align}
In this case Eq. (\ref{eq:energyrelation}) for the final energy $\varepsilon_{f}$
becomes

\begin{align}
\begin{split}
\varepsilon_{f}^{2}&=\left(2n+1\right)\sqrt{e\kappa(p_{x,i}+\varepsilon_{i}-\omega(1+\text{cos}\theta))}\\
&+(p_{x,i}-\omega\text{cos}\theta)^{2}+\omega^{2}\text{sin}^{2}\theta\text{sin}^{2}\Phi+m^{2}.\label{eq:effunccos}
\end{split}
\end{align}
Now we wish to carry out the integration over $d\text{cos}\theta$
so we must transform the delta function so

\begin{equation}
\delta(\varepsilon_{f}(\text{cos}\theta)+\omega-\varepsilon_{i})=\frac{\delta(\text{cos}\theta-\text{cos}\theta_{0})}{|\frac{d\varepsilon_{f}}{d\text{cos}\theta}(\text{cos}\theta)|},
\end{equation}
where $\text{cos}\theta_{0}$ is the solution to the equation 
\begin{equation}
\varepsilon_{f}(\text{cos}\theta)+\omega-\varepsilon_{i}=0,\label{eq:efzero}
\end{equation}
which we will find later. We obtain from Eq. (\ref{eq:effunccos})

\begin{align}
\begin{split}
2\varepsilon_{f}\frac{d\varepsilon_{f}}{d\text{cos}\theta}&=\left(2n+1\right)\frac{-e\kappa\omega}{2\sqrt{e\kappa(p_{x,i}
+\varepsilon_{i}-\omega(1+\text{cos}\theta))}}\\
&+2(p_{x,i}-\omega\text{cos}\theta)(-\omega)\\
&+\omega^{2}\text{sin}^{2}\Phi\frac{d}{d\text{cos}\theta}(\text{sin}^{2}\theta),
\end{split}
\end{align}
from which we can isolate $\frac{d\varepsilon_{f}}{d\text{cos}\theta}$.

\begin{align}
\begin{split}
\frac{d\varepsilon_{f}}{d\text{cos}\theta}=\frac{1}{\varepsilon_{f}}&\left(-\left(2n+1\right)\frac{e\kappa\omega}{4\sqrt{e\kappa(p_{x,i}+\varepsilon_{i}-\omega(1+\text{cos}\theta))}}\right.\\
&\left.-\omega p_{x,i}+\omega^{2}\text{cos}\theta\text{cos}^{2}\Phi\right).
\end{split}
\end{align}
And so we have integrated over all the delta functions and can write the differential rate as

\begin{equation}
dW=\sum_{n_{f}}\frac{e^{2}}{8\pi\omega'}\left|\int\bar{U}_{f}(y)\slashed{\varepsilon}^{*}U_{i}(y)e^{-ik_{y}y}dy\right|^{2}\frac{1}{|\frac{d\varepsilon_{f}}{d\text{cos}\theta}|}\omega{}^{2}d\omega d\Phi.\label{eq:finalrate}
\end{equation}
To find the solution of Eq. (\ref{eq:efzero}) we will recall that
we consider ultra-relativistic particles such that $\theta$ is small
meaning we can perform the series expansions of $\text{cos}\theta\simeq1-\frac{\theta^{2}}{2}$
and $\text{sin}\theta\simeq\theta$. Inserting this in Eq. (\ref{eq:effunccos})
we obtain

\begin{align}
\begin{split}
\left(\varepsilon_{i}-\omega\right)^{2}&=\left(2n+1\right)\sqrt{e\kappa(p_{x,i}+\varepsilon_{i}-\omega(2-\frac{\theta^{2}}{2})}\\
&+(p_{x,i}-\omega+\omega\frac{\theta^{2}}{2})^{2}+\omega^{2}\theta^{2}\text{sin}^{2}\Phi+m^{2},
\end{split}
\end{align}
which leads us to the sought after solution of Eq. (\ref{eq:efzero})

\begin{align}
\begin{split}
\theta_{0}&=\left[\left(\varepsilon_{i}-\omega\right)^{2}-\left(2n+1\right)\sqrt{e\kappa(p_{x,i}+\varepsilon_{i}-2\omega)}\right.\\
&\left.-(p_{x,i}-\omega)^{2}-m^{2}\right]^{1/2}\\
&/\left[(p_{x,i}-\omega)\omega+\omega^{2}\text{sin}^{2}\Phi\right.\\
&\left.+\frac{\omega/4}{p_{x,i}+\varepsilon_{i}-2\omega}\left(2n+1\right)\sqrt{e\kappa(p_{x,i}+\varepsilon_{i}-2\omega)}\right]^{1/2}.\label{eq:thetaeq}
\end{split}
\end{align}
Now we have all the quantities necessary to evaluate the rate from
Eq. (\ref{eq:finalrate}).

\section{Baier method\label{sec:Baier-method}}

From \cite{Wistisen_2014} it can be seen that the differential power
emitted in the semi-classical operator method is given by

\begin{align}
\begin{split}
\frac{d^{2}P}{d\omega d\Omega}&=\frac{1}{T}\frac{e^{2}}{4\pi^{2}}\omega'^{2}\left(\frac{\varepsilon^{2}+\varepsilon'^{2}}{2\varepsilon^{2}}\left|\intop_{-\infty}^{\infty}\left(\boldsymbol{n}-\boldsymbol{v}\right)e^{i\omega'(t-\boldsymbol{n}\cdot\boldsymbol{r})}dt\right|^{2}\right.\\
&\left.+\frac{\omega^{2}m^{2}}{2\varepsilon^{4}}\left|\intop_{-\infty}^{\infty}e^{i\omega'(t-\boldsymbol{n}\cdot\boldsymbol{r})}dt\right|^{2}\right),\label{eq:baierrad}
\end{split}
\end{align}
where $\varepsilon'=\varepsilon-\omega$,
$\omega'=\omega\varepsilon/(\varepsilon-\omega)$, $\boldsymbol{n}=(\text{cos}\theta,\text{sin}\theta\text{cos}\Phi,\text{sin}\theta\text{sin}\Phi)$
is the direction of emission and $\boldsymbol{r}(t)$ and $\boldsymbol{v}(t)$
are the classical position and velocity vectors, respectively. It is beneficial to first look at the integral from the second term
and insert the motion found in section \ref{sec:Classical-motion}

\begin{align}
\begin{split}
&\intop_{-\infty}^{\infty}e^{i\omega'(t-\boldsymbol{n}\cdot\boldsymbol{r})}dt\\
=&\intop_{-\infty}^{\infty}e^{i\omega'(t-\text{cos}\theta\left[\left(1-\frac{1}{2\gamma_0^{2}}-\frac{\xi^{2}}{4\gamma_0^{2}}\right)t-\frac{1}{4}\left(\frac{\xi}{\gamma_{0}}\right)^{2}\frac{\text{sin}(2\omega_{0}t)}{2\omega_{0}}\right])}\\
&\times e^{-i\omega'\text{sin}\theta\text{cos}\Phi\frac{\xi}{\gamma_{0}\omega_{0}}\text{sin}(\omega_{0}t)}dt.
\end{split}
\end{align}
If we change variable to $\tau=\omega_{0}t$ and expand $\text{cos}\theta$
as earlier this can be rewritten as

\begin{align}
\begin{split}
&\intop_{-\infty}^{\infty}e^{i\omega'(t-\boldsymbol{n}\cdot\boldsymbol{r})}dt\\
=&\frac{1}{\omega_{0}}\intop e^{i\omega'(\frac{\tau}{2\gamma_0^{2}\omega_{0}}\left[1+\frac{1}{2}\xi^{2}+\gamma_{0}^{2}\theta^{2}\right])}\\
&\times e^{i\omega'\left(\frac{1}{8\omega_{0}}\left(\frac{\xi}{\gamma_{0}}\right)^{2}\text{sin}(2\tau)-\text{sin}\theta\text{cos}\Phi\frac{\xi}{\gamma_{0}\omega_{0}}\text{sin}(\tau)\right)}d\tau.\label{eq:nosum}
\end{split}
\end{align}
Now we know that

\begin{equation}
e^{i\omega'\left(\frac{1}{8\omega_{0}}\left(\frac{\xi}{\gamma_{0}}\right)^{2}\text{sin}(2\tau)-\text{sin}\theta\text{cos}\Phi\frac{\xi}{\gamma_{0}\omega_{0}}\text{sin}(\tau)\right)},
\end{equation}
is a $2\pi$ periodic function so we can write it as a Fourier series

\begin{equation}
e^{i\omega'\left(\frac{1}{8\omega_{0}}\left(\frac{\xi}{\gamma_{0}}\right)^{2}\text{sin}(2\tau)-\text{sin}\theta\text{cos}\Phi\frac{\xi}{\gamma_{0}\omega_{0}}\text{sin}(\tau)\right)}=\sum_{n=-\infty}^{\infty}c_{n}e^{-in\tau},
\end{equation}
with coefficients given by

\begin{align}
\begin{split}
c_{n}&=\frac{1}{2\pi}\int_{-\pi}^{\pi}e^{i\left(n\tau+\frac{\omega'}{8\omega_{0}}\left(\frac{\xi}{\gamma_{0}}\right)^{2}\text{sin}(2\tau)-\text{sin}\theta\text{cos}\Phi\frac{\omega'\xi}{\gamma_{0}\omega_{0}}\text{sin}(\tau)\right)}\\
&=A_{0}(-n,-\alpha_{1},-\alpha_{2})=A_{0}(n,\alpha_{1},\alpha_{2}),
\end{split}
\end{align}
where we have defined

\begin{align}
\begin{split}
&A_{m}(n,\alpha_{1},\alpha_{2})\\
&=\frac{1}{2\pi}\int_{-\pi}^{\pi}\text{cos}^{m}(\tau)e^{i\left(\alpha_{1}\text{sin}(\tau)-\alpha_{2}\text{sin}(2\tau)-n\tau\right)}d\tau,
\end{split}
\end{align}
as in \cite{Ritus_1985,Baie98}, with

\begin{equation}
\alpha_{1}=\text{sin}\theta\text{cos}\Phi\frac{\omega'\xi}{\gamma_{0}\omega_{0}},\label{eq:alfapar}
\end{equation}
and

\begin{equation}
\alpha_{2}=\frac{\omega'}{8\omega_{0}}\left(\frac{\xi}{\gamma_{0}}\right)^{2}.\label{eq:betapar}
\end{equation}
When inserting this in Eq. (\ref{eq:nosum}) we obtain

\begin{align}
\begin{split}
&\intop_{-\infty}^{\infty}e^{i\omega'(t-\boldsymbol{n}\cdot\boldsymbol{r})}dt\\
&=\frac{1}{\omega_{0}}\sum_{n=-\infty}^{\infty}c_{n}\intop e^{i\tau(\frac{\omega'}{2\gamma_0^{2}\omega_{0}}\left[1+\frac{1}{2}\xi^{2}+\gamma_{0}^{2}\theta^{2}\right]-n)}d\tau\\
&=\frac{2\pi}{\omega_{0}}\sum_{n=-\infty}^{\infty}A_{0}(n,\alpha_{1},\alpha_{2})\delta(\frac{\omega'}{2\gamma_0^{2}\omega_{0}}\left(1+\frac{1}{2}\xi^{2}+\gamma_{0}^{2}\theta^{2}\right)-n).\label{eq:Jterm}
\end{split}
\end{align}
Now calculating $\intop_{-\infty}^{\infty}\left(\boldsymbol{n}-\boldsymbol{v}\right)e^{i\omega'(t-\boldsymbol{n}\cdot\boldsymbol{r})}dt$
is straightforward. For the $y$-component we have

\begin{align}
\begin{split}
&\intop_{-\infty}^{\infty}\left(\boldsymbol{n}-\boldsymbol{v}\right)_{y}e^{i\omega'(t-\boldsymbol{n}\cdot\boldsymbol{r})}dt\\
&=\intop_{-\infty}^{\infty}\left(\text{sin}\theta\text{cos}\Phi-\frac{\xi}{\gamma_{0}}\text{cos}(\omega_{0}t)\right)\\
&\times e^{i\omega'(t-\text{cos}\theta v_{0}t-\text{sin}\theta\text{cos}\Phi\frac{\xi}{\gamma_{0}\omega_{0}}\text{sin}(\omega_{0}t))}dt.
\end{split}
\end{align}
The first term is simply a constant (no $\tau$ dependence) times
the integral we have already calculated, and the cosine factor in
the second term means we simply need to replace $A_{0}(n,\alpha_{1},\alpha_{2})$
with $A_{1}(n,\alpha_{1},\alpha_{2})$ and so

\begin{align}
\begin{split}
&\intop_{-\infty}^{\infty}\left(\boldsymbol{n}-\boldsymbol{v}\right)_{y}e^{i\omega'(t-\boldsymbol{n}\cdot\boldsymbol{r})}dt\\
&=\frac{2\pi}{\omega_{0}}\sum_{n=-\infty}^{\infty}\left[\text{sin}\theta\text{cos}\Phi A_{0}(n,\alpha_{1},\alpha_{2})-\frac{\xi}{\gamma_{0}}A_{1}(n,\alpha_{1},\alpha_{2})\right]\\
&\times\delta(\frac{\omega'}{2\gamma_0^{2}\omega_{0}}\left(1+\frac{1}{2}\xi^{2}+\gamma_{0}^{2}\theta^{2}\right)-n).
\end{split}
\end{align}
The $z$-component is simply a constant times the result from
Eq. (\ref{eq:Jterm}), since $v_{z}=0$, so we have

\begin{align}
\begin{split}
&\intop_{-\infty}^{\infty}\left(\boldsymbol{n}-\boldsymbol{v}\right)_{z}e^{i\omega'(t-\boldsymbol{n}\cdot\boldsymbol{r})}dt\\
&=\frac{2\pi}{\omega_{0}}\sum_{n=-\infty}^{\infty}\text{sin}\theta\text{sin}\Phi A_{0}(n,\alpha_{1},\alpha_{2})\\
&\times\delta(\frac{\omega'}{2\gamma_0^{2}\omega_{0}}\left(1+\frac{1}{2}\xi^{2}+\gamma_{0}^{2}\theta^{2}\right)-n).
\end{split}
\end{align}
In Eq. (\ref{eq:baierrad}) we see that we need these quantities squared
and so we must consider the meaning of the delta function $\delta(\frac{\omega'}{2\gamma_0^{2}\omega_{0}}\left(1+\frac{1}{2}\xi^{2}+\gamma_{0}^{2}\theta^{2}\right)-n)$
squared. This came from an integral over the phase $\tau$ and so
the usual approach is that 
\begin{align}
\begin{split}
\left[\delta(\frac{\omega'}{2\gamma_0^{2}\omega_{0}}\left(1+\frac{1}{2}\xi^{2}+\gamma_{0}^{2}\theta^{2}\right)-n)\right]^{2}\\
=\delta(\frac{\omega'}{2\gamma_0^{2}\omega_{0}}\left(1+\frac{1}{2}\xi^{2}+\gamma_{0}^{2}\theta^{2}\right)-n)\frac{\Delta\tau}{2\pi},\label{eq:deltasq}
\end{split}
\end{align}
where $\Delta\tau$ is the phase-length which is $\omega_{0}T$ where
$T$ is the interaction time, which we can divide with on both sides
of Eq. (\ref{eq:baierrad}) to obtain the energy emitted per unit
time. We therefore obtain that

\begin{align}
\begin{split}
&\left|\intop\left(\boldsymbol{n}-\boldsymbol{v}\right)e^{i\omega'(t-\boldsymbol{n}\cdot\boldsymbol{r})}dt\right|^{2}\\
&=\frac{(2\pi)^{2}}{\omega_{0}^{2}}\frac{T\omega_{0}}{2\pi}\delta(\frac{\omega'}{2\gamma_0^{2}\omega_{0}}\left(1+\frac{1}{2}\xi^{2}+\gamma_{0}^{2}\theta^{2}\right)-n)\\
&\times\sum_{n=-\infty}^{\infty}\left[\text{sin}\theta\text{cos}\Phi A_{0}(n,\alpha_{1},\alpha_{2})-\frac{\xi}{\gamma_{0}}A_{1}(n,\alpha_{1},\alpha_{2})\right]^{2}\\
&+\left[\text{sin}\theta\text{sin}\Phi A_{0}(n,\alpha_{1},\alpha_{2})\right]^{2}.
\end{split}
\end{align}
Now we wish to carry out the integration over $d\theta$ so we recognize
that the content of the delta function is a function of $\theta$
which we define as $g(\theta)$ to use the formula $\delta(g(\theta))=\delta(\theta-\theta_{0,\text{B}})/|\frac{dg}{d\theta}(\theta_{0,\text{B}})|$
where $\theta_{0,\text{B}}$ is the solution to $g(\theta_{0,\text{B}})=0$. So

\begin{equation}
g(\theta)=\frac{\omega'}{2\gamma_0^{2}\omega_{0}}\left(1+\frac{1}{2}\xi^{2}+\gamma_{0}^{2}\theta^{2}\right)-n,\label{eq:f}
\end{equation}

\begin{equation}
g'(\theta)=\frac{\omega'}{\omega_{0}}\theta,\label{eq:fprime}
\end{equation}
and we then find the zero as

\begin{equation}
\theta_{0,\text{B}}=\frac{1}{\gamma_{0}}\sqrt{\frac{2\gamma_0^{2}\omega_{0}n}{\omega'}-\left(1+\frac{\xi^{2}}{2}\right)}.\label{eq:thetaemit}
\end{equation}
The other solution, $-\theta_{0,\text{B}}$ is not allowed by our choice of coordinate system where $0\leq\theta\leq\pi$. Here we see that we should have $n\geq1$ to have any solutions. Once
again approximating $\text{sin}\theta\simeq\theta$ we have that

\begin{align}
\begin{split}
&\int d\Phi\theta d\theta\left|\intop\left(\boldsymbol{n}-\boldsymbol{v}\right)e^{i\omega'(t-\boldsymbol{n}\cdot\boldsymbol{r})}dt\right|^{2}\\
&=\int d\Phi\frac{2\pi T}{\omega'}\sum_{n=-\infty}^{\infty}\left[\theta_{0,\text{B}}\text{cos}\Phi A_{0}(n,\alpha_{1},\alpha_{2})-\frac{\xi}{\gamma_{0}}A_{1}(n,\alpha_{1},\alpha_{2})\right]^{2}\\
&+\left[\theta_{0,\text{B}}\text{sin}\Phi A_{0}(n,\alpha_{1},\alpha_{2})\right]^{2},
\end{split}
\end{align}
and for the second term of Eq. (\ref{eq:baierrad}), we obtain

\begin{align}
\begin{split}
&\left|\intop e^{i\omega'(t-\boldsymbol{n}\cdot\boldsymbol{r})}dt\right|^{2}=\frac{2\pi}{\omega_{0}^{2}}T\omega_{0}\sum_{n=-\infty}^{\infty}A_{0}^{2}(n,\alpha_{1},\alpha_{2})\\
&\times\delta(\frac{\omega'}{2\gamma_0^{2}\omega_{0}}\left(1+\frac{1}{2}\xi^{2}+\gamma_{0}^{2}\theta^{2}\right)-n).
\end{split}
\end{align}
Integrating this term over all angles as well, we obtain
\begin{align}
\begin{split}
&\int d\Phi\theta d\theta\left|\intop e^{i\omega'(t-\boldsymbol{n}\cdot\boldsymbol{r})}dt\right|^{2}\\
&=\int d\Phi\frac{2\pi T}{\omega'}\sum_{n=-\infty}^{\infty}A_{0}^{2}(n,\alpha_{1},\alpha_{2}).
\end{split}
\end{align}
So in total we obtain the emitted power $dP$ (energy per unit time) differential in the emitted photon energy as

\begin{figure}[t]
\includegraphics[width=1\columnwidth]{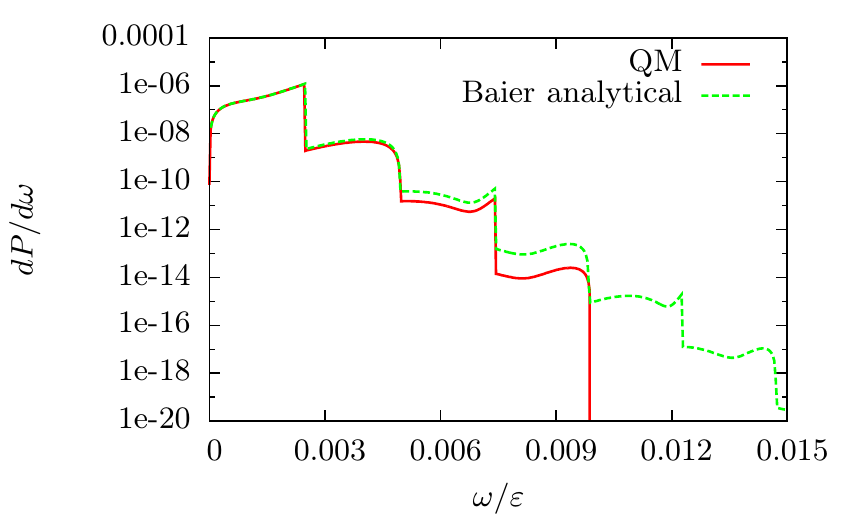}

\caption{The case of $n=4$ and $\xi=0.1$. The label 'QM' refers to the fully quantum calculation and 'Baier analytical' to the semi-classical method of Baier \& Katkov.\label{fig:dipol}}
\end{figure}

\begin{align}
\begin{split}
&\frac{dP}{d\omega}=\frac{e^{2}}{2\pi}\omega'\int d\Phi\sum_{n=1}^{\infty}\\
&\times\left(\frac{\varepsilon'^{2}+\varepsilon^{2}}{2\varepsilon^{2}}\left[\left\{ \theta_{0,\text{B}}\text{cos}\Phi A_{0}(n,\alpha_{1},\alpha_{2})-\frac{\xi}{\gamma_{0}}A_{1}(n,\alpha_{1},\alpha_{2})\right\} ^{2}\right.\right.\\
&\left.+\left\{ \theta_{0,\text{B}}\text{sin}\Phi A_{0}(n,\alpha_{1},\alpha_{2})\right\} ^{2}\right]\\
&\left.+\frac{\omega^{2}m^{2}}{2\varepsilon^{4}}A_{0}^{2}(n,\alpha_{1},\alpha_{2})\right).\label{eq:baierradfinal1}
\end{split}
\end{align}
In this form it is clear which terms correspond to which from Eq.
(\ref{eq:baierrad}), however it is not immediately obvious that it
is identical to that found in \cite{Baie98}. To obtain this we must
carry out the square in the term $\left\{ \theta_{0,\text{B}}\text{cos}\Phi A_{0}(n,\alpha_{1},\alpha_{2})-\frac{\xi}{\gamma_{0}}A_{1}(n,\alpha_{1},\alpha_{2})\right\} ^{2}$.
This will give us a a term $-2\theta_{0,\text{B}}\text{cos}\Phi A_{0}(n,\alpha_{1},\alpha_{2})\frac{\xi}{\gamma_{0}}A_{1}(n,\alpha_{1},\alpha_{2})$
which we will rewrite by employing the relation found in \cite{Ritus_1985}
stating that

\begin{align}
\begin{split}
\alpha_{1}A_{1}(n,\alpha_{1},\alpha_{2})&=(n-2\alpha_{2})A_{0}(n,\alpha_{1},\alpha_{2})\\
&+4\alpha_{2}A_{2}(n,\alpha_{1},\alpha_{2}),\label{eq:identity}
\end{split}
\end{align}
and from this we can express $\text{cos}\Phi A_{1}(n,\alpha_{1},\alpha_{2})$
in terms of $A_{0}$ and $A_{2}$, which after some rewriting will
lead us to the expected result

\begin{align}
\begin{split}
&\frac{dP}{d\omega}=\frac{e^{2}}{2\pi}\frac{\omega}{\gamma_{0}^{2}}\int d\Phi\sum_{n=1}^{\infty}\\
&\times\left(-A_{0}^{2}+\xi^{2}\left(1+\frac{u^{2}}{2\left(1+u\right)}\right)\left(A_{1}^{2}-A_{0}A_{2}\right)\right),\label{eq:baierosc}
\end{split}
\end{align}
where $u=\frac{\omega}{\varepsilon-\omega}$.

\section{Results and discussion\label{sec:discuss}}

Now we have calculated the radiation emission using two different
approaches to the same problem. One is fully quantum mechanical, and
the other a semi-classical approach. A third method which is well
known in the literature \cite{PhysRevLett.111.054802,PhysRevLett.112.015001,PhysRevLett.105.220403,Wistisen2018} is to approximate the emission as happening
in a constant crossed field and use the formula for radiation emission
in this case. This is generally considered applicable when $\xi\gg1$
and so we will also make this comparison when this condition is fulfilled.

When dealing with radiation emission the quantum parameter is defined
by

\begin{figure}[t]
\includegraphics[width=1\columnwidth]{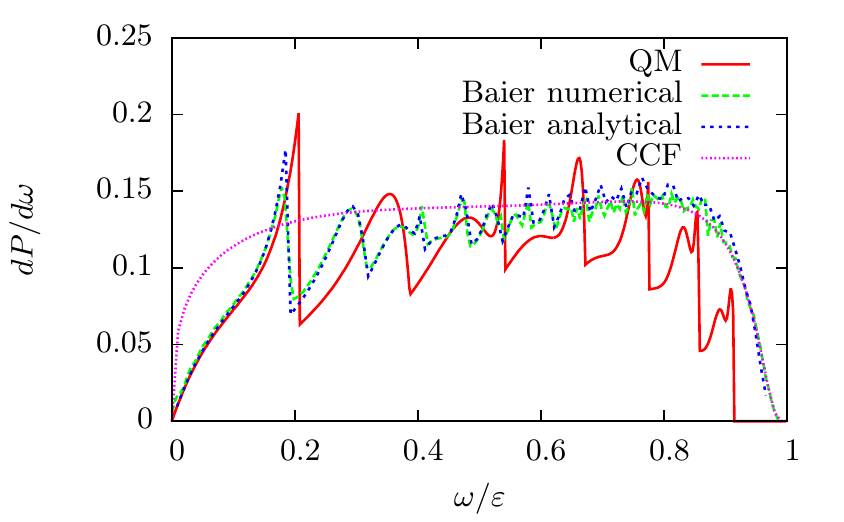}\caption{The case of $n=8$, $\xi=5$, $\chi=7.81$. The label 'QM' refers to the fully quantum calculation, 'Baier numerical' to the semi-classical method of Baier \& Katkov for 15 periods of oscillation, 'Baier analytical' to the analytical results obtained from the semi-classical method corresponding to the limit of many oscillations, and the label 'CCF' corresponds to the constant crossed field approximation. This figure shows the 'doubly quantum' regime where the semi-classical method breaks down.\label{fig:quantumcase}}
\end{figure}

\begin{equation}
\chi=\frac{e\sqrt{-(F^{\mu\nu}p_{\nu})^{2}}}{m^{3}},\label{eq:chidef}
\end{equation}
and tells us how important quantum effects such as recoil and spin
are. When $\chi\ll1$ these effects are small. However the effect
of low quantum number is not covered by this parameter, and is a separate
condition. We will consider the peak value in terms of the parameters
of our problem which is then given by

\begin{equation}
\chi_{\text{max}}=\kappa y_{\text{max}}\frac{2\gamma_0}{E_{c}},\label{eq:chimax}
\end{equation}
where $E_{c}=\frac{m^{2}}{e}$ is the Schwinger critical field. From
the wave-functions seen in Eq. (\ref{eq:spinup}) we know that the
harmonic oscillator function $I_{n}(\eta)$ will start tending towards
$0$ when $\eta>\sqrt{2n}$, corresponding to the classical amplitude
of oscillation. Thus $y_\text{max}$ can also be accurately written as $y_{\text{max}}=L\sqrt{2n}$ when $n$ is large so

\begin{equation}
\chi_{\text{max}}=\kappa\sqrt{2n}L\frac{2\gamma_0}{E_{c}}.
\end{equation}
Setting $p_{x}\simeq\varepsilon$ and $\beta_\text{b} \simeq 1$ we have

\begin{align}
&\frac{1}{L}\simeq\sqrt[4]{2e\kappa p_{x}},\\
&\frac{1}{2p_{x}L^{4}}=e\kappa\label{eq:kappaquant}
\end{align}
and so we have that

\begin{equation}
\chi_{\text{max}}=\sqrt{2n}\left(\frac{\lambda_{C}}{L}\right)^{3},\label{eq:chimax-1}
\end{equation}
where $\lambda_{C}=1/m$ is the Compton wavelength. In the special case of the harmonic oscillator, the momentum space wave-function is the same as the space wave function save only for a different variable, such that instead of $I_{n}(y/L)$ we have $I_{n}(q_y L)$ and therefore since the function $I_n(\eta)$ decreases rapidly for $\eta^2 > 2n$ we can also write $q_{\text{y,max}}^2=2n/L^2$.
And so we express
the other parameter usually considered when dealing with radiation
emission $\xi$ in terms of the parameters of our solutions to obtain
that

\begin{figure}[t]
\includegraphics[width=1\columnwidth]{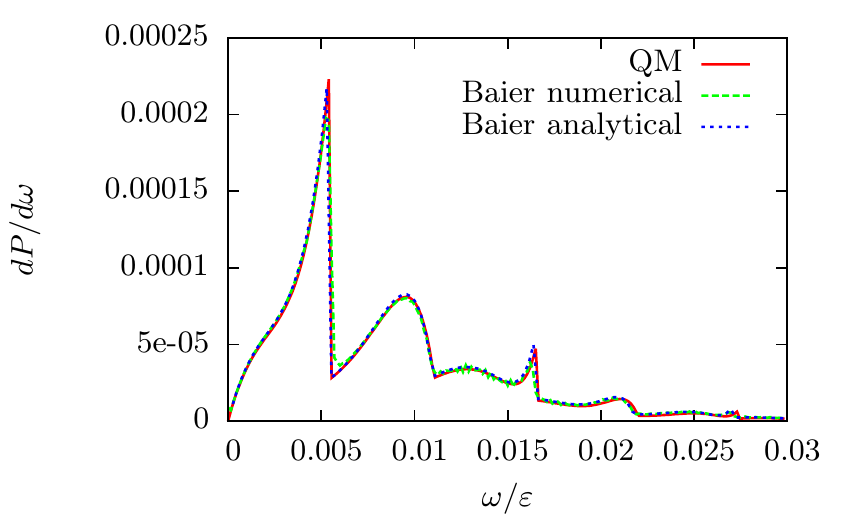}

\caption{The case of $n=120$, $\xi=1$. The label 'QM' refers to the fully quantum calculation, 'Baier numerical' to the semi-classical method of Baier \& Katkov for 15 periods of oscillation, 'Baier analytical' to the analytical results obtained from the semi-classical method corresponding to the limit of many oscillations. Here we see how for quite large values of the quantum number the semi-classical method is good.\label{fig:intn120}}
\end{figure}

\begin{equation}
\xi=\frac{q_{\text{y,max}}}{m}=\frac{\lambda_{C}}{L}\sqrt{2n}.\label{eq:zetaN}
\end{equation}
And combining this with Eq. (\ref{eq:chimax-1}) we obtain the useful relation

\begin{equation}
\chi_{\text{max}}=\frac{\xi^{3}}{2n}.
\end{equation}
For weak fields, meaning a small field gradient $\kappa$, $L$ will
become large, and so we will denote $L\gg\lambda_{C}$ as the weak
field gradient regime and vice versa. Below we will discuss features
of the radiation spectrum shown in the figures in the different regimes.
In the figures the label 'QM' corresponds to the exact calculation
of Eq. (\ref{eq:finalrate}), 'Baier analytical' corresponds to Eq.
(\ref{eq:baierosc}). 'Baier numerical' corresponds to using the formula
of Eq. (\ref{eq:baierrad}) by numerically solving the equations of
motion corresponding to a time of $15$ oscillations in the field
numerically, and then performing the integration over angles and time
numerically as done in \cite{Wistisen_2014}. The label 'CCF'
corresponds to the radiation emitted when applying the constant crossed
field approximation.

\subsection{Weak field gradient regime, $L\gg\lambda_{C}$}

In this regime when the quantum number $n$ is small, both $\chi_{\text{max}}$
and $\xi$ will be small as seen from Eq. (\ref{eq:chimax-1}) and
Eq. (\ref{eq:zetaN}). A small value of $\chi_{\text{max}}$ means
the only quantum effects for $n$ small are those due to the quantization
of the motion. Since $\xi$ will be small, the radiation is in the
dipole regime, meaning different harmonics are clear and most radiation
comes from the first harmonic. In \figref{dipol} we have shown a
plot of the radiation spectrum in this regime using the exact calculation
and the semi-classical approximation. Coincidentally the two calculations
yield same result for the first harmonic, and differences are only
seen for higher harmonics. So differences are only seen in the
parts of the spectrum where the radiation yield is small. Another
difference is that the exact calculation only allows a finite number
of harmonics corresponding to transitions from the initial state with
quantum number $n_{i}$ to one with lower quantum number, and the
last harmonic thus corresponds to transition to the ground state and for photon energies 
above the threshold corresponding to this harmonic, no radiation can
be emitted. With the semi-classical method this is not the case and the
sum over harmonics is infinite. This case of the weak field gradient
regime is the regime of planar channeling of positrons for low energies.
For channeling of relativistic positrons, the positron also experiences
a potential which is close to parabolic. However in channeling, low
quantum numbers for the motion can only be achieved while one is also
in the dipole regime. Therefore in \cite{Baie98} there is a section
on calculating the (dominant) radiation from the first harmonic in
this regime of low quantum numbers which does not use their developed
semi-classical method. Although planar channeling of positrons is
in this regime, we think that it would be difficult to see the effects
seen in \figref{dipol} because the slight anharmonicity of the true
potential would also yield contributions at higher harmonics, which
would likely be larger than these effects.

\begin{figure}[t]
\includegraphics[width=1\columnwidth]{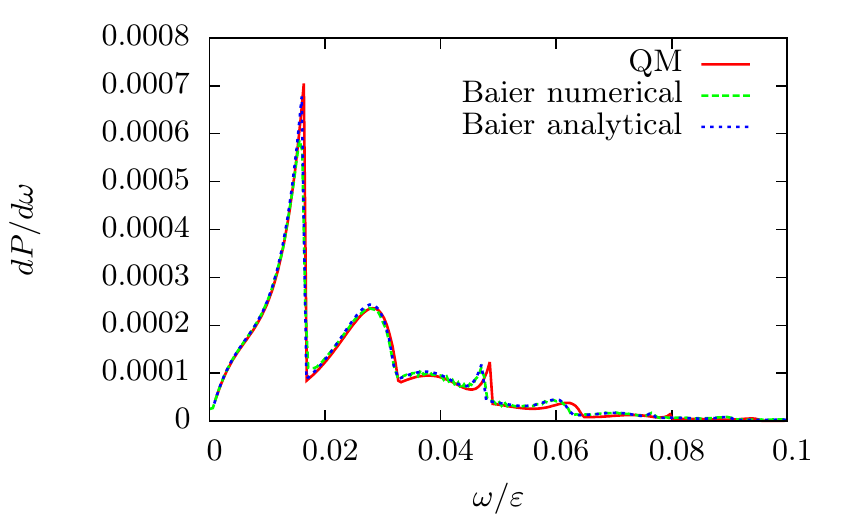}\caption{The case of $n=40$, $\xi=1$. The label 'QM' refers to the fully quantum calculation, 'Baier numerical' to the semi-classical method of Baier \& Katkov for 15 periods of oscillation, 'Baier analytical' to the analytical results obtained from the semi-classical method corresponding to the limit of many oscillations. Here we see how for smaller values of the quantum number one begins to see small deviations between the semi-classical method and the correct result.\label{fig:intn40}}
\end{figure}

\subsection{Strong field gradient regime, $L\ll\lambda_{C}$}

If one wants to see big differences in the whole of the spectrum as
seen in \figref{quantumcase}, one must be in the strong field regime
and have small value of $n$. In this regime one will always have
large $\chi$ such that recoil and spin is important, and for small
$n$ one has the additional quantum effect of the quantized motion
i.e. one needs the wave-function instead of the trajectory.

In \figref{quantumcase} we have shown an example of this which could
be dubbed the ``doubly quantum regime'', where it is seen that the
correct calculation deviates significantly from the constant crossed
field approximation (CCF) but also from the result of the semi-classical
operator method which is more general, but evidently fails in the
regime of low quantum numbers. This transition from the usual regime
to the doubly quantum regime happens when the $n$ quantum number is small
and $L=\frac{1}{m}$ which can be related to a certain beam density
as we will see below.

\begin{figure}[t]
\includegraphics[width=1\columnwidth]{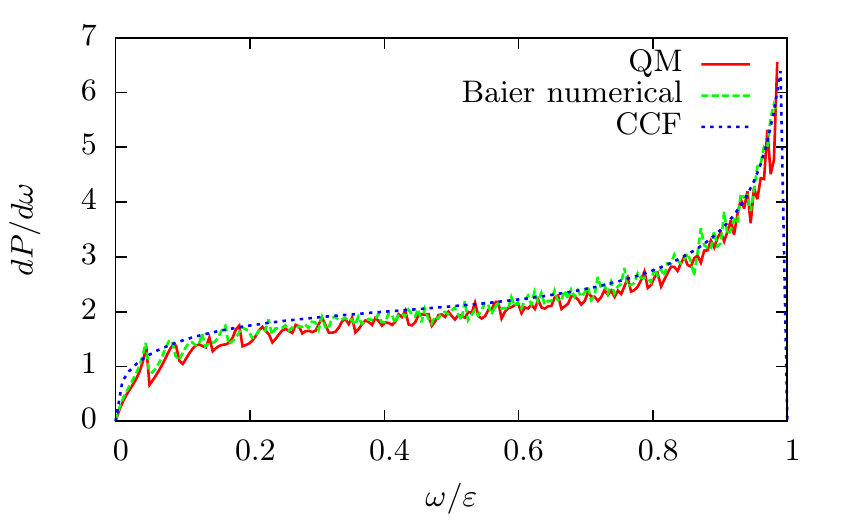}\caption{The case of $n=40$, $\xi=30$. The label 'QM' refers to the fully quantum calculation, 'Baier numerical' to the semi-classical method of Baier \& Katkov for 15 periods of oscillation and 'CCF' to the constant crossed field approximation. Here we see how, also in the regime where the CCF approximation may be applied, the deviation from the correct results is small when $n=40$.\label{fig:CCFregime}}
\end{figure}

In \figref{intn120} and \figref{intn40} we show the radiation spectrum
for $\xi=1$ but for a different value of the quantum number $n$
of the radiating particles. Here it is seen that as $n$ is large
as in the $n=120$ case the agreement between the exact calculation
and the semi-classical approach is good while when it becomes smaller,
in the case of $n=40$ the agreement becomes worse. In \figref{CCFregime}
we show the radiation spectrum in the regime where the constant crossed
field approximation is applicable and see that while there are small
differences between the semi-classical method and the exact calculation
in the position of the harmonics, see e.g. the position of 3rd and
4th harmonic peak, the overall size of the spectrum coincides quite
well, while the CCF approximation seems to slightly overestimate the
radiation emitted at low frequencies. So to see major differences
one needs an even smaller value of $n$ as is the case seen in \figref{quantumcase}.

\subsection{Beam parameter considerations}

To gain an understanding of when these new effects could arise in
beamstrahlung, we wish to approximate the parameters we have introduced
in terms of the usually given beam parameters, see \tabref{Beam-parameters.}.
From Eq. (\ref{eq:kappa}) we have that $\kappa$ relates to the peak
density as

\begin{equation}
\kappa=4\pi\rho_{0},\label{eq:kappadens}
\end{equation}
where $\rho_{0}=\rho(0,0,0)$. So using Eq. (\ref{eq:kappaquant})
we have that

\begin{equation}
\frac{1}{L^{4}}=8\pi e\rho_{0}\varepsilon.
\end{equation}
Now we can define the critical density $\rho_{c}$ as that corresponding
to $L=1/m$ so $m^{4}=8\pi e\rho_{c}\varepsilon$, or

\begin{equation}
\rho_{c}=\frac{m^{4}}{8\pi e\varepsilon}.\label{eq:Criticaldens}
\end{equation}
So to be in the doubly quantum regime one should reach this density
and have a small quantum number, i.e. that the transverse beam size
becomes comparable to the Compton wavelength. This also gives us another
way of expressing the length parameter of the problem $L$ in terms
of the critical density since

\begin{equation}
\frac{1}{L^{4}}=m^{4}\frac{\rho_{0}}{\rho_{c}},\label{eq:Lpar}
\end{equation}
and so we also have that

\begin{table*}[t]
\begin{tabular}{|c|c|c|c|c|c|c|}
\hline 
Machine & CLIC & CLIC & ILC & ILC & HER 2017  & CLIC mod.\tabularnewline
\hline 
\hline 
$\varepsilon$ & 190 GeV & 1500 GeV & 100 GeV & 250 GeV & 4 GeV & 1500 GeV\tabularnewline
\hline 
N & $5.2\times10^{9}$ & $3.7\times10^{9}$ & $2.0\times10^{10}$ & $2.0\times10^{10}$ & $6.5\times10^{10}$ & $3.7\times10^{9}$\tabularnewline
\hline 
$\Sigma_{x}$ & 149nm & 40nm & 904nm & 474nm & 10.7$\mu$m & 4$\mu$m\tabularnewline
\hline 
$\Sigma_{y}$ & 2.9nm & 1nm & 7.8nm & 5.9nm & 62nm & 10pm\tabularnewline
\hline 
$\Sigma_{z}$ & 70$\mu$m & 44$\mu$m & 300$\mu$m & 300$\mu$m & 5mm & 44$\mu$m\tabularnewline
\hline 
$\chi_{\text{max}}$ & $0.32$ & $10.7$ & $0.025$ & $0.12$ & $1.65\times10^{-5}$ & 0.11\tabularnewline
\hline 
$\rho_{\text{max}}$ & $7.13\times10^{6}\text{eV}^{3}$ & $8.72\times10^{7}\text{eV}^{3}$ & $3.92\times10^{5}\text{eV}^{3}$ & $9.89\times10^{5}\text{eV}^{3}$ & $8.16\times10^{2}\text{eV}^{3}$ & $8.72\times10^{7}\text{eV}^{3}$\tabularnewline
\hline 
$\rho_{c}$ & $1.67\times10^{11}\text{eV}^{3}$ & $2.12\times10^{10}\text{eV}^{3}$ & $3.18\times10^{11}\text{eV}^{3}$ & $1.27\times10^{11}\text{eV}^{3}$ & $7.9\times10^{12}\text{eV}^{3}$ & $2.12\times10^{10}\text{eV}^{3}$\tabularnewline
\hline 
$\lambda_{C}/L$ & 0.081 & 0.25 & 0.033 & 0.0528 & 0.0032 & 0.25\tabularnewline
\hline 
$\sqrt{2n_{\text{max}}}$ & 608 & 657 & 674 & 808 & 512 & 6.6\tabularnewline
\hline 
$\xi_{\text{max}}$ & 49 & 166 & 22.5 & 43 & $1.63$ & 1.67\tabularnewline
\hline 
\end{tabular}
\caption{Beam parameters.\label{tab:Beam-parameters.}}
\end{table*}

\begin{equation}
\left(\frac{\lambda_{C}}{L}\right)^{4}=\frac{\rho_{0}}{\rho_{c}}.
\end{equation}
Now we can use Eq. (\ref{eq:chimax-1}) to obtain an estimate of the
quantum number corresponding to the particles with the largest amplitude,
which will contribute most to the radiation spectrum. This gives us

\begin{equation}
\sqrt{2n_{\text{max}}}=\chi_{\text{max}}/\left(\frac{\rho_{0}}{\rho_{c}}\right)^{3/4}.
\end{equation}
Now we wish to obtain an expression giving us the quantum number corresponding
to the largest amplitude of oscillation when crossing a bunch in terms
of the usual beam parameters such that we can see how this scales
and in what regime these effects would become important. We have that

\begin{align}
\begin{split}
\frac{\rho_{0}}{\rho_{c}}&=\frac{Ne}{(2\pi)^{3/2}\Sigma_{x}\Sigma_{y}\Sigma_{z}}\frac{8\pi e\varepsilon}{m^{4}}\\
&=\frac{8\pi Ne^{2}\gamma_{0}}{(2\pi)^{3/2}\Sigma_{x}\Sigma_{y}\Sigma_{z}m^{3}}.
\end{split}
\end{align}
Now we use Eq. (\ref{eq:chimax}) and insert $\kappa$ from Eq. (\ref{eq:kappa})
and set $y_{\text{max}}=\Sigma_{y}$

\[
\chi_{\text{max}}=\frac{4\gamma_0 Ne^{2}}{\sqrt{2\pi}m^{2}\Sigma_{x}\Sigma_{z}},
\]
So introducing $o_{i}=\Sigma_{i}m$ we have

\begin{align}
\begin{split}
\chi_{\text{max}}/\left(\frac{\rho_{0}}{\rho_{c}}\right)^{3/4}&=\frac{4\gamma_0 Ne^{2}}{\sqrt{2\pi}o_{x}o_{z}}\left(\frac{(2\pi)^{3/2}o_{x}o_{y}o_{z}}{8\pi Ne^{2}\gamma_{0}}\right)^{3/4}\\
&=\sqrt[4]{\frac{4}{\sqrt{2\pi}}}\frac{(o_{x}o_{y}o_{z})^{3/4}}{o_{x}o_{z}}(N\gamma_0 e^{2})^{1/4},
\end{split}
\end{align}
and so

\begin{equation}
n_{\text{max}}=\frac{1}{\sqrt[4]{2\pi}}\frac{(o_{x}o_{y}o_{z})^{3/2}}{(o_{x}o_{z})^{2}}(N\gamma_0 e^{2})^{1/2}.\label{eq:nmax}
\end{equation}
To obtain an expression for $\xi$ we can use the expression for the
amplitude Eq. (\ref{eq:amplitude}) and set it equal $\Sigma_{y}$
so that

\begin{equation}
\xi_{\text{max}}=\Sigma_{y}\gamma_{0}\sqrt{\frac{2e\kappa}{\varepsilon}}=\frac{2}{(2\pi)^{1/4}}\sqrt{\frac{Ne^{2}\gamma_{0} o_{y}}{o_{x}o_{z}}}.
\end{equation}
From the fact that $y_{\text{max}}=L\sqrt{2n}$ and that the transition
to the doubly quantum regime happens when $L\simeq1/m$ and $n$ small,
we can get an estimate of how small the beam-size has to be. So we
have that $\Sigma_{y,\text{crit}}\simeq10\lambda_{C}\sqrt{20}=17$
pm. Currently the accelerator SuperKEKB has beams with a size of 62
nm while future machines such as CLIC has proposed 1 nm beams. In
\tabref{Beam-parameters.} we have shown the beam parameters of a
current electron-positron accelerator, superKEKB, along with some
that are still on the drawing board namely the CLIC and ILC. From
the large value of $\sqrt{2n_{\text{max}}}$ it is clear that the
effects we have seen in this paper will not be important, i.e. the
semi-classical method will provide the correct result. If the beams
are reshaped, making them even smaller in the $y$-direction and larger
in the $x$-direction, $n_{\text{max}}$ will go down, as can be seen
from Eq. (\ref{eq:nmax}). For HER (superKEKB) if we reduce $\Sigma_{y}$
by a factor of $100$ and increase $\Sigma_{x}$ with the same amount,
the luminosity is unchanged but we then have $n_{\text{max}}=13$.
However, because of the low energy and so, not being close to the
critical density $\rho_{c}$ we would only be in the weak field gradient
regime where $\chi_{\text{max}}\ll1$. To be in the doubly quantum
regime we have considered CLIC with 3 TeV center of mass energy and
reshaped the beams in the same way. This is the ``CLIC mod.'' case
from \tabref{Beam-parameters.}. Here it is seen that $\chi_{\text{max}}=0.11$,
so the usual quantum effects would start to come into play, and at
the same time we have a small quantum number of $n_{\text{max}}=22$.
This reshaping of the beams would be beneficial since $\chi$ is reduced
while keeping the luminosity the same, thus reducing the emitted energy
to beamstrahlung. This is the original purpose of having the bunches
shaped like sheets. We see here, that if this strategy is taken to
the extreme, one will enter this new regime of radiation emission.

\section{Conclusion\label{sec:Conclusion}}

From first principles we have calculated the radiation emission from
a relativistic spin-$\frac{1}{2}$ particle in a harmonic oscillator
like potential which showed interesting features, allowing us to find
a new regime of radiation emission where another quantum effect besides
the usual ones, come into play, namely the quantization of the motion.
This effect is absent in the well studied examples of non-linear Compton
scattering in a plane wave where the semi-classical method of Baier
and Katkov yields the correct result. Contrary, in the field configuration
studied here, the semi-classical operator method fails for low quantum
numbers. This is interesting in itself, but we also tried to see when
one would enter this regime in the case of beamstrahlung. We found
that current machines and the ones currently on the drawing board
are far away from being in this regime, however if the strategy of
making bunches shaped like sheets, which is the strategy to avoid
energy loss due to beamstrahlung, is taken to the extreme, one enters
this new regime of radiation emission where the quantization of the
motion of the radiating particle becomes important.

\section{Acknowledgments}
This work was partially supported by a research grant (VKR023371) from VILLUM FONDEN.

\bibliographystyle{unsrt}

\end{document}